\documentclass[letterpaper,12pt]{article}
\RequirePackage{amsmath,amssymb,amsfonts,amsbsy,amsthm,booktabs,epsfig,graphicx,rotating, geometry}
\usepackage[labelfont = bf]{caption}
\usepackage{booktabs}
\usepackage{tcolorbox}
\usepackage{comment}
\usepackage{authblk}
\usepackage{float}

\usepackage[natbibapa,nodoi]{apacite}
\newcommand{\beginsupplement}{%
        \setcounter{table}{0}
        \renewcommand{\thetable}{S\arabic{table}}%
        \setcounter{figure}{0}
        \renewcommand{\thefigure}{S\arabic{figure}}%
        \setcounter{page}{1}
        \renewcommand{\thepage}{S\arabic{page}}
        \setcounter{section}{0}
        \renewcommand{\thesection}{S\arabic{section}}%
     }

\begin{document}

\title{Where Does Haydn End and Mozart Begin?\\Composer Classification of String Quartets}

\author[a]{Katherine C. Kempfert}
\author[b]{Samuel W.K. Wong}
\affil[a]{Department of Statistics, University of California Berkeley}
\affil[b]{Department of Statistics and Actuarial Science, University of Waterloo}
\date{}

\maketitle

\begin{abstract}
For centuries, the history and music of Joseph Franz Haydn and Wolfgang Amadeus Mozart have been compared by scholars. Recently, the growing field of music information retrieval (MIR) has offered quantitative analyses to complement traditional qualitative analyses of these composers. 
In this MIR study, we classify the composer of Haydn and Mozart string quartets based on the content of their scores. 
Our contribution is an interpretable statistical and machine learning approach that provides high classification accuracies and musical relevance.
We develop novel global features that are automatically computed from symbolic data and informed by musicological Haydn-Mozart comparative studies, particularly relating to the sonata form. Several of these proposed features are found to be important for distinguishing between Haydn and Mozart string quartets. Our Bayesian logistic regression model attains leave-one-out classification accuracies over 84\%, higher than prior works and providing interpretations that could aid in
assessing musicological claims.
Overall, our work can
help expand the longstanding dialogue surrounding Haydn and Mozart and exemplify the benefit of interpretable machine learning in MIR, with potential applications to music generation and classification of other classical composers.
\end{abstract}

\section{Introduction}

Music information retrieval (MIR) is an interdisciplinary field that has grown as digitalised music data and computing power have become widely available. Methods have been developed to automatically perform many types of tasks in MIR: composer, genre, and mood classification \citep{pollastri2001classification}, \citep{tzanetakis2002musical}, \citep{laurier2008multimodal}; query, such as matching a sung melody to a song \citep{kosugi2000practical}; generation of novel music \citep{johanson1998gp}; and recommender systems for consumers, such as on Spotify and Pandora \citep{van2013deep}. Thus, MIR has become increasingly relevant to how music is both studied and enjoyed.  For a review of MIR and its applications, see \citep{downie2003music} and \citep{schedl2014music}.

For the \textit{composer classification} task in MIR, probabilistic models are used to predict the composer of music as one among several classes. Quantitative results from such models can expand on traditional qualitative analyses of composers in musicology. In particular, these models can potentially detect patterns in the music that would be unrecognizable to humans and provide validation of subjective musical interpretations via predictive performance. In addition to improving musicological understanding of composers, findings from composer classification can impact MIR tasks including music generation, recommendation systems, and authorship attribution. For example, \cite{backer2005musical} performed composer classification of 30 fugues from J.S. Bach, W.F. Bach and J.L. Krebs; based on the results from decision trees and other classifiers, \citeauthor{backer2005musical} identified important musical features that separate these composers and determined J.L. Krebs to be the most likely composer of a fugue with disputed authorship. 
The identified features
could additionally be used as constraints for music generation in J.S. Bach's style, a challenging task dating back to 1988 \citep{ebciouglu1988expert} that has more recently motivated the development of generation algorithms based on machine learning such as DeepBach \citep{hadjeres2017deepbach}.

In this MIR study, we focus on two stylistically similar classical composers writing within the same genre and time period. We develop a statistical machine learning approach to classify Haydn and Mozart string quartets based on the content of their scores, motivated by the historical and cultural significance of these composers, the difficulty of the task, and the applications. 

\bigskip
\textbf{Significance of Haydn and Mozart as composers.} For centuries, the music and history of Haydn and Mozart have been compared by scholars. According to Robert L. Marshall, `The critical and scholarly literature devoted to this repertoire is nothing short of oceanic and includes contributions from some of the most profound musical thinkers of the past two centuries--among them such authorities as Hermann Abert, Friedrich Blume, Wilhelm Fischer, Leonard Ratner, Charles Rosen, and Donald Francis Tovey' \cite[Preface]{harutunian2005haydn}. More recent comparative analyses include \textit{Metric Manipulations in Haydn and Mozart} \citep{mirka2009metric} and \textit{Haydn's and Mozart's Sonata Styles: A Comparison} \citep{harutunian2005haydn}. Mirka argues that Haydn's music is `artful popularity', `appealing to all kinds of listeners', while Mozart's `overwhelming art', stemming from `harmonic and polyphonic complexity $\hdots$ required greater intellectual involvement of listeners $\hdots$' (p. 303). 
Harutunian confirms the overwhelming artistry of Mozart, repeatedly referring to his music as `operatic' (p. 65, 81) and even citing this as a reason for his greater success over Haydn in the opera. These differences between Haydn and Mozart are only a few simple examples of the many complex qualitative comparisons undertaken over the centuries. 

However, Haydn and Mozart had many similarities: `They were not only contemporaneous composers, using the harmonic vocabulary of the late eighteenth century at a time when its syntax was the most restricted and defined, but they shared the summit in the development of $\hdots$ the sonata style' \cite[Foreword]{harutunian2005haydn}. At times, members of royalty commissioned both Haydn and Mozart (for example, King Frederick William II of Prussia), which may have further constrained Mozart's and Haydn's compositions to be similar \citep{zaslaw1990compleat}. The two composers had similar patrons and cultural upbringings, both Austrians active in Vienna during periods of their lives \citep{zaslaw1990compleat}. In addition to their shared cultural influences, the composers directly influenced each other, with `quartet playing $\hdots$ central to contact between Haydn and Mozart' \cite[p. 54]{larsen1997new}. In fact, Mozart dedicated his Op. 10 set of six string quartets to Haydn. After hearing a performance of the quartets, Haydn told Mozart's father Leopold, `I tell you before God as an honest man that your son is the greatest composer known to me either in person or by name. He has taste, and what is more, the most profound knowledge of composition' \citep[p. 264]{zaslaw1990compleat}.

\bigskip
\textbf{Difficulty of the classification task.}  While we have noted some musicological distinctions between Haydn's and Mozart's compositions, many listeners of their music fail to hear any differences, possibly due to the more conspicuous similarities. Listeners' difficulties in distinguishing between Haydn and Mozart string quartets can be exemplified by the results of an informal online quiz (created by Craig Sapp and Yi-Wen Liu of Stanford University and accessed at \texttt{http://qq.themefinder.org}). The user is prompted to answer a series of questions (including number of years in classical music training, instruments one can play, and familiarity with Haydn and Mozart), then to identify randomly selected Haydn and Mozart string quartets. Even the users with maximal music experience have not achieved more than 67\% accuracy on average. Although this quiz is not a random and representative survey, the results still evidence the difficulty of the Haydn-Mozart string quartet classification task.

Prior MIR studies on Haydn-Mozart string quartets have yielded composer classification accuracies between 70\% and 80.5\%, exceeding the human accuracy of 67\% (see Table \ref{table:lit_acc}). However, these accuracies are substantially lower than attained from classifying some other sets of stylistically similar composers. For example, \cite{van2005musical} and \cite{hontanilla2013modeling} highlighted that works by Handel and Telemann could be classified with optimum accuracies 12\% to 18\% higher than for Haydn-Mozart string quartets. Additionally, \cite{hontanilla2013modeling} found that fugues by J.S. Bach and fugues by Shostakovich (written in homage to J.S. Bach) are more distinguishable than Haydn-Mozart string quartets, reporting an 18.6 percentage point difference in optimum classification accuracies.

\bigskip

\textbf{Applications.}   We mention several applications for the Haydn-Mozart classification task.
First, musically interpretable features identified as important for distinguishing between the two composers may be useful for evaluating musicological claims, such as that Mozart's compositions are more `operatic' or `dramatic' than Haydn's (Sections 3.2.2 and 5.4). 
Second, one can envision incorporating
these features within algorithms to generate synthetic music in the style of Mozart, Haydn, or their combination that matches the feature constraints, e.g., see \cite{briot2020deep} for a review of music generation algorithms based on deep learning.  Third, features and methods that work well for discriminating between Haydn and Mozart might be generalised for music analysis and classification of other classical composers.

\bigskip

\textbf{Contribution of this work.} Our contribution in this study is a statistical machine learning approach leading to high classification accuracies, model-based interpretations, and insights relevant to musicologists. As argued by \cite{murdoch2019interpretable}, interpretability in the context of statistical machine learning should be defined by three principles: `predictive accuracy, descriptive accuracy, and relevancy, with relevancy judged relative to a human audience' (p. 1). 
As in many other prior MIR studies, our classification models include \textit{global features}, namely variables extracted from the musical scores of Haydn and Mozart string quartets, e.g., summary statistics such as the mean and standard deviation of pitch in the cello voice. We additionally develop and interpret novel features inspired by the sonata structure of Mozart and Haydn compositions. While these features are informed by musicological Haydn-Mozart comparative studies, they are automatically computed from scores and require no expert structural analysis or hand-coding; as such, they could be potentially extended to classify other classical composers, particularly from the Classical period or writing in sonata or ternary form. 
Our classification models reach a leave-one-out cross-validation (LOO CV) accuracy of about 84\%, exceeding the previous benchmark and offering interpretability. 
Therefore, our work suggests that inclusion of  
features involving structure could improve composer classification of stylistically similar composers in the symbolic domain. In addition to the potential applications of this work to music analysis, classification, and generation, we believe our work exemplifies the benefit of interpretable machine learning in MIR.

In the next section, we present our dataset. The motivation and calculation of musicologically inspired features are discussed in Section 3. In Section 4, we detail our statistical machine learning methods, including feature selection and the classification model for discriminating between Haydn and Mozart string quartets. 
The results are presented, compared to prior studies, and musically interpreted in Section 5. We conclude our paper and suggest further directions of research in Section 6. Finally, the datasets and source code for our methods are publicly available at \texttt{https://github.com/wongswk/haydn-mozart}. 

\section{Data}
Music data can be expressed in the form of auditory or symbolic information. Audio formats include live performances and recordings, such as MP3 files, CDs and tapes, while symbolic formats include scores, text, and computer encodings like Musical Instrument Digital Interface (MIDI) and **kern \citep{downie2003music}. Auditory formats can be affected by a certain performance or performer’s interpretation of a written score, which in turn can vary to some extent based on print edition. Symbolic formats transcribe the written score directly, which nonetheless may still depend on the print edition and means of transcription. Our goal is to understand Haydn and Mozart as composers, particularly the main aspects of their music that would be heard by listeners, read by performers, and discussed by musicologists. Hence, a symbolic format is preferred, from which we parse the basic pitch and duration information that would be largely invariant across musical representations of the compositions.
To our knowledge (as of writing in May 2019), all other Haydn-Mozart classification studies have also used symbolic formats, including MIDI (\citealp{kaliakatsos2011weighted,herlands2014machine,hontanilla2013modeling, velarde2018convolution}) and **kern (\citealp{van2005musical, hillewaere2010string,taminau2010applying, velarde2016composer, velarde2018convolution}). 

We opt to use the **kern symbolic format of music, as it is the most commonly used in the aforementioned prior studies. Our proposed features, including those informed by musicological comparative studies, are thus calculated using pitch and duration (including barlines) information from the **kern scores. However, we note that our proposed features do not explicitly rely on **kern encoding and could be flexibly applied to MIDI or other symbolic formats.
A discussion of **kern and other symbolic formats can be found in \citep{selfridge1997beyond} and  \citep{huron2002music}.

We use two datasets containing **kern representations of Haydn and Mozart string quartet scores. Each string quartet has one to five movements, with each movement containing the four standard voices (or parts), 
Violin 1, Violin 2, Viola, and Cello. The Mozart scores additionally include two flute quartets and one oboe quartet, which have been treated as part of the string quartet genre in prior studies (\citealp{van2005musical, hillewaere2010string,taminau2010applying, velarde2016composer, velarde2018convolution}); they contain a woodwind voice instead of a violin voice. 
We refer to these datasets as HM285 and HM107 and describe them as follows.
\begin{itemize}
    \item \textbf{HM285.} Our main dataset is obtained from the KernScores website \newline (\texttt{http://kern.humdrum.org/}), which is maintained by the Center for Computer Assisted Research in the Humanities (CCARH) at Stanford University. There are 82 Mozart string quartet movements and 210 Haydn string quartet movements available on the website, representing the majority of known string quartet movements by these composers: 86 movements authored by Mozart and 280 by Haydn. There are 7 **kern files with errors in the encoding of scores, so we omit the corresponding movements from our analysis. Thus, HM285 consists of 82 Mozart movements and 203 Haydn movements. 
    \item \textbf{HM107.} To facilitate comparison with previous studies, we additionally apply our methods to the HM107 dataset, originally obtained from the CCARH KernScores website above and used by \cite{van2005musical} and \cite{velarde2016composer, velarde2018convolution}. It consists of 54 Haydn movements and 53 Mozart movements.
\end{itemize}

\begin{figure}
\includegraphics[width=1\textwidth]{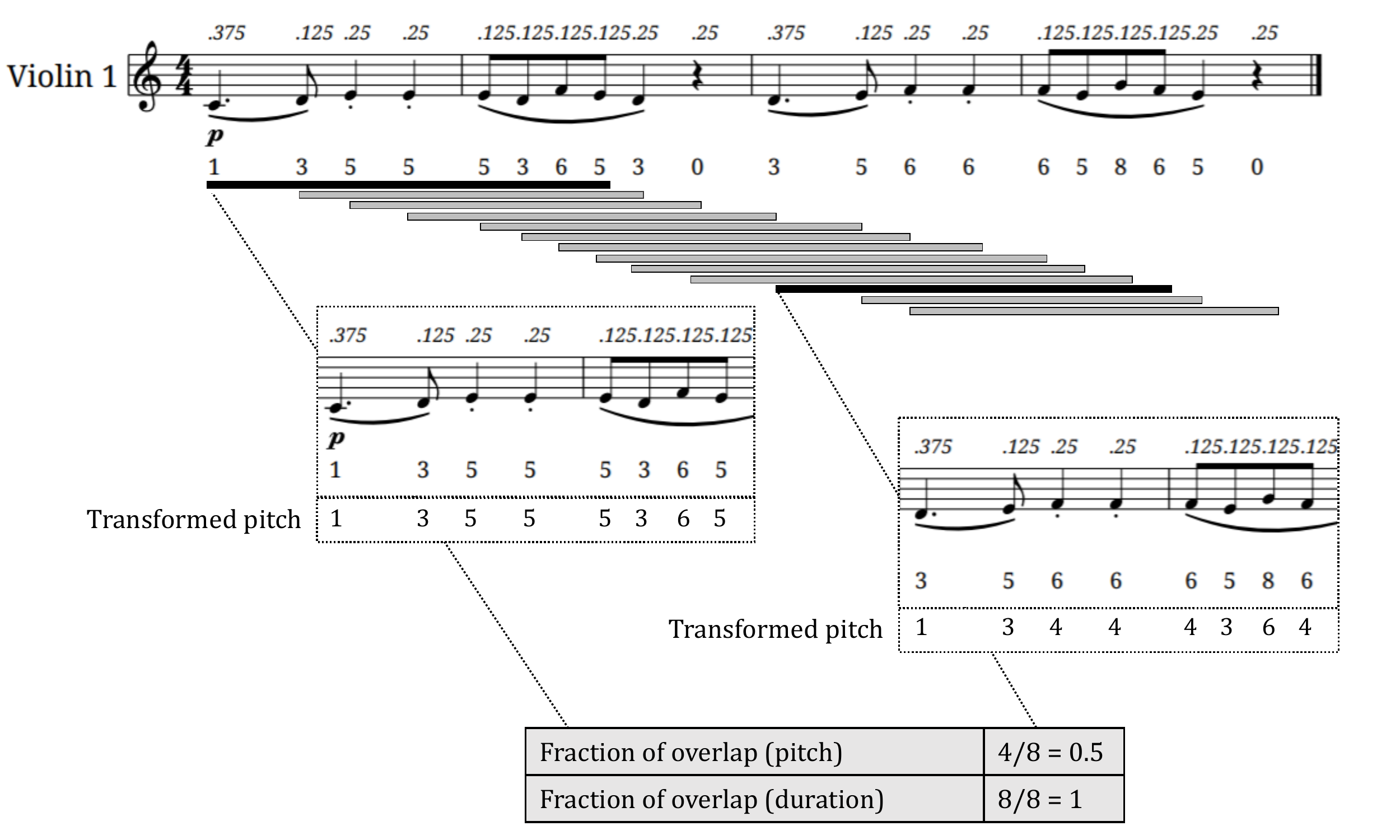}
\centering
\caption{Encoding of an excerpt from Mozart's String Quartet No. 4, Mvmt. 1, in C Major (K. 157). (Top) Encoded pitch values and duration values are displayed below and above the original score, respectively. (Bottom) We show an example of how sonata-inspired segment features (Section 3.2) are calculated for segments of length $m = 8$ notes. The pitches are transformed to start on 1 in each segment, so the pitches in segment 1 and segment 11 are transformed as 1, 3, 5, 5, 5, 3, 6, 5 and 1, 3, 4, 4, 4, 3, 6, 4, respectively. Segments 1 and 11 have the same encoded duration values of 0.375, 0.125, 0.25, 0.25, 0.125, 0.125, 0.125, 0.125. As a result, these two segments' fraction of overlap for $m = 8$ is $4/8 = 0.5$ for pitch and $8/8 = 1$ for duration.}
\label{fig:encoding}
\end{figure}

We process the data in the statistical programming environment R \citep{rdevelopment2008r}. For each voice in each movement, pitch and duration information are parsed 
from the **kern files. Hence, each movement is represented with eight tracks: pitch and duration tracks for all four voices. The proposed features in Section 3 will be subsequently computed from these eight tracks.
As an example, Figure \ref{fig:encoding} (top) displays our pitch and duration encodings for several bars of the Violin 1 part of Mozart's string quartet No.4, movement 1 (K. 157), as we now describe.

Each voice generally only plays one note at a time, such as seen in Figure \ref{fig:encoding}. Chords and harmonic intervals in a single voice (known as \emph{multiple stopping}) occur very infrequently, so for simplicity we retain only the highest of simultaneous notes in those cases. Rests are encoded as 0. The pitch of each note is encoded as an integer between 1 and 12 (except in 3.2.2 when intervals are calculated), following the order of the chromatic scale: $1,2,3,\ldots, 12$ corresponds to C, C\#, D, $\ldots$, B respectively. Thus, we retain pitch-class information; for example, middle C is encoded as 1, as are any higher or lower Cs. This representation is commonly used in genre classification \citep{correa2016survey} and facilitates analysis by capturing the most meaningful aspect of pitch according to the distributional view of key-finding, which posits that listeners perceive key via the distribution of pitch-classes in a piece \citep{temperley2008pitch}.

The duration of each note is encoded as the fraction of time it makes up in a bar. For example, in common time, a quarter note is encoded as $0.25$. Therefore, the length of the bar is implicitly encoded in the parsed duration information.

\section{Feature Development and Extraction}

Feature development involves proposing a litany of global features that may help to discriminate between Haydn and Mozart string quartets. Feature extraction refers to our algorithmic methods for automatically computing these features from the data representation introduced in Section 2. The novelty in our approach to feature development and extraction is in quantifying the qualitative differences that have been discussed at length in scholarly Haydn-Mozart comparisons, particularly through our \textit{sonata-inspired} features.
A concise subset of the most important features for classification models will be subsequently selected by statistical methods, as discussed in Section 4. Therefore, we can gain insights from both selected and unselected features: selected features suggest areas in which Haydn and Mozart string quartets differ, while unselected features might point to similarities between the composers.

\subsection{Review of the Sonata Form}
In the qualitative analysis \textit{Haydn's and Mozart's Sonata Styles: A Comparison}, musicologist John Harutunian (2005) states, `Central to the music of Haydn and Mozart is the concept of sonata style' (p. 1). Hence, we propose features inspired by the sonata form. Our musically informative yet flexible \textit{sonata-inspired} features are intended to balance interpretability, predictability, and generalizability of the resulting statistical models. Comparably to pattern discovery in MIR, the features are automatically computed using concepts of similarity and variability among notes in sections of a piece. An overview of pattern discovery and structure analysis in MIR can be found in \citep{ong2006structural}.

We note that our sonata-inspired features are calculated algorithmically, so they can be applied to any piece without requiring expert hand-coding. As long as a piece roughly follows ternary form (i.e., an A-B-A structure), they are expected to extract relevant
musical information. The sonata is the most common structural form of Haydn and Mozart string quartet movements. Though not all movements strictly follow the sonata form, they often contain similar structure. For example, movements in the rondo form follow the pattern A-B-A-C-A-B-A (or a variation) and thus have similar structural elements. Hence, we believe meaningful information can be gleaned from these sonata-inspired features in our datasets.

As the sonata form is essential to understanding these features, we provide a brief summary based on \cite[p. 1-2]{harutunian2005haydn}. 
A piece of music in sonata form has three sections: the exposition, development, and recapitulation.
\begin{enumerate}
\item In the \textbf{exposition}, the basic thematic material of the sonata is presented. The beginning key is known as the tonic. As the exposition ends, the key modulates, so that it generally ends in a different key from which it started. 
\item In the \textbf{development}, one or more themes from the exposition are altered, and some new material may be introduced. The development often contains the greatest amount of change.
\item In the \textbf{recapitulation}, the opening material is revisited, but it is all in the home key, giving a `sense of resolution and completion' \cite[p. 1]{harutunian2005haydn}. In general, the recapitulation begins with the opening material in the tonic.
\end{enumerate}

\subsection{Feature Categories}
This section presents the list of quantitative features that we compute for each Haydn and Mozart string quartet movement, along with descriptions of their musical significance. 
We incorporate musicological knowledge from Haydn-Mozart comparative studies, in particular the aspects of sonata form discussed in \citep{harutunian2005haydn}.
We complete our feature set by including some that have worked well in previous studies. We may organise our features into five main categories:  
basic summary, interval, exposition, development, and recapitulation. These features are summarised in Table \ref{table:feature} and discussed in depth in the following subsections, 3.2.1 to 3.2.5. The remainder of this subsection defines and describes additional terminology used therein. 

\textbf{\textit{Monophonic and Polyphonic Features.}}
As appropriate to each category, \textit{monophonic} and \textit{polyphonic} features are considered. Monophonic features are intended to measure the specific melodic and rhythmic role of each separate voice, while polyphonic features can capture the interaction between voices.

\textbf{\textit{Segment Features.}}
Within each category, many of our proposed features can be further described as higher-order
\textit{segment} features, since they utilise segments (or sliding windows) of notes.
Let $M$ denote the total number of notes in a voice of a movement and $m$ the desired \textit{segment length}, i.e., number of notes in the segment. 
Then for a voice of a movement, a segment feature is calculated over $M - m + 1$ segments of note positions: $\{1, 2,\hdots, m\}$, $\{2, 3, \hdots, m + 1\}$, ..., $\{M - m + 1, M - m + 2, \hdots,  M\}$.
The \textit{location} of a segment is the order of the segment divided by the total number of segments, e.g., the location of segment $\{2, 3, \hdots, m + 1\}$ above is $2/(M - m + 1)$.
Segment features can be applied to pitch or duration tracks, as appropriate.
As an example, Figure \ref{fig:encoding} (bottom) shows the 13 segments of length $m = 8$ notes that would be calculated from $M = 20$ notes in the Violin 1 part of Mozart's string quartet No. 4, movement 1 (truncated to 4 bars for illustration).

\textbf{\textit{Lengths of Segment Features.}}  The segment length of $m$ notes will control the amount of musical context for the segment features. We compute all segment features for a range of reasonable candidates for $m$: $8,10,12,14,16,18$.
These candidates are intended as a compromise between reducing the number of computed features while capturing musical context over a range of lengths from 8 to 18 notes, based on our exploratory data analyses that are included in the Supplementary Material.
Recall that the most important features for classification will  subsequently be identified by feature selection, as discussed in Section 4.

\textbf{\textit{Transforming Pitch for Sonata-Inspired Segment Features.}} 
For sonata-inspired segment features (Sections 3.2.3-3.2.5) calculated from pitch tracks, we perform an additional transformation of pitches. For each segment of length $m$ notes within the voice of a movement, the segment's pitches are transformed relative to the segment's first note. As demonstrated in Figure \ref{fig:encoding}, the segment of pitches D, E, F, F, F, E, G, F is first encoded as 3, 5, 6, 6, 6, 5, 8, 6 then transformed to 1, 3, 4, 4, 4, 3, 6, 4. A segment with the same relative pitches, such as E, F$\sharp$, G, G, G, F$\sharp$, A, G, would result in the same transformation of 1, 3, 4, 4, 4, 3, 6, 4. Since most listeners perceive pitch relatively, rather than in terms of absolute frequency \citep{levitin2005absolute}, this transformation enables detection of musical phrases that sound the same to most listeners, even if the phrases are in different keys.

\textbf{\textit{Fraction of Overlap for Sonata-Inspired Segment Features.}} The exposition and recapitulation features (Sections 3.2.3 and 3.2.5) include \textit{fraction of overlap} and \textit{fraction of overlap count at threshold $t$} features. 
For duration or pitch, the \textit{fraction of overlap} is computed as the proportion of notes that match in two segments of notes. For the example with $m = 8$ in Figure \ref{fig:encoding}, the two segments of pitches C, D, E, E, E, D, F, E and D, E, F, F, F, E, G, F have a fraction of overlap equaling $4/8 = 0.5$, since their transformed pitches 1, 3, 5, 5, 5, 3, 6, 5 and 1, 3, 4, 4, 4, 3, 6, 4 are the same for 4 out of 8 note positions. Each of the two segments have the encoded duration values of 0.375, 0.125, 0.25, 0.25, 0.125, 0.125, 0.125, 0.12, so their fraction of overlap for duration is $8/8 = 1$. The \textit{fraction of overlap count at threshold $t$} is the number of segment pairs with a fraction of overlap at or above $t$. The thresholds $t$ we use for calculating these features are discussed in Sections 3.2.3 and 3.2.5.

\begin{table*}
    \caption{Features for Mozart and Haydn String Quartet Scores} 
    \label{table:feature}
    \resizebox{\linewidth}{!}{
	\begin{tabular}{@{}lll@{}}
    \toprule
\textbf{Feature Category} &\textbf{Duration} &\textbf{Pitch}\\
\midrule
\textbf{Basic Summary} & &Number of notes$^{M, N}$\\
 (22 features) &Mean and standard deviation of duration$^{M, N}$ &Mean and standard deviation of pitch$^{M, N}$\\
 & &Proportion of simultaneous rests$^{P, N}$\\
 & &Proportion of simultaneous notes$^{P, N}$\\
\midrule
\textbf{Interval} & &Proportion of each pairwise interval class$^{M, N}$\\
(392 features) & &\hspace{1cm}Voicepair differences in proportion of pairwise interval classes$^{P, N}$\\
& &Proportion of each pairwise interval mode$^{M, N}$\\
& &Proportion of each pairwise interval sign$^{M, N}$\\
&Mean and standard deviation of pairwise interval differences$^{M, N}$ &Mean and standard deviation of pairwise interval semitones$^{M, N}$\\
& &\hspace{1cm}Voicepair differences of mean and standard deviation of pairwise interval semitones$^{P, N}$\\
& &Summary statistics for proportion of minor third relative pitch intervals within a segment\\
& &\hspace{1cm}Minimum, first quartile, median, third quartile, maximum$^{M, S}$\\
& &\hspace{1cm}Mean and standard deviation$^{M, S}$\\
& &\hspace{1cm}Count of segments with proportion $0$ and at or above $0.6$$^{M, S}$\\
\midrule
\textbf{Exposition} &Maximum fraction of overlap with opening material within first half of movement$^{M, S}$ &Maximum fraction of overlap with opening material within first half of movement$^{M, S}$\\
(240 features) &\hspace{1cm}Percentile of maximum fraction of overlap match$^{M, S}$ &\hspace{1cm}Percentile of maximum fraction of overlap match$^{M, S}$\\
&\hspace{1cm}Fraction of overlap counts at thresholds 0.7, 0.9, and 1$^{M, S}$ &\hspace{1cm}Fraction of overlap counts at thresholds 0.7, 0.9, and 1$^{M, S}$\\
\midrule
\textbf{Development} &Maximum standard deviation over all segments of fixed length$^{M, S}$ &Maximum standard deviation over all segments of fixed length$^{M, S}$\\
(288 features) &\hspace{1cm}Percentile of maximum standard deviation segment$^{M, S}$ &\hspace{1cm}Percentile of maximum standard deviation segment$^{M, S}$\\
&\hspace{1cm}Count of standard deviations at thresholds 0.7, 0.8, 0.9, and 0.95$^{M, S}$  &\hspace{1cm}Count of standard deviations at thresholds 0.7, 0.8, 0.9, and 0.95$^{M, S}$\\
\midrule
\textbf{Recapitulation} &Maximum fraction of overlap with opening material$^{M, S}$&Maximum fraction of overlap with opening material$^{M, S}$\\
(240 features) &\hspace{1cm}Percentile of maximum fraction of overlap$^{M, S}$&\hspace{1cm}Percentile of maximum fraction of overlap$^{M, S}$\\
&\hspace{1cm}Fraction of overlap counts at thresholds 0.7, 0.9, and 1$^{M, S}$ &\hspace{1cm}Fraction of overlap counts at thresholds 0.7, 0.9, and 1$^{M, S}$ \\
\bottomrule
   \end{tabular}
   }
   \scriptsize
   Our novel sonata-inspired features are 
   those in the Exposition, Development, and Recapitulation feature categories.
   The superscripts $M$, $P$, $S$, $N$ indicate features that are monophonic, polyphonic, segment, and non-segment, respectively. Monophonic features are calculated for 4 voices, and segment features are calculated for 6 different segment lengths of 8, 10, 12, 14, 16, 18 notes. In total, there are $22 + 392 + 240 + 288 + 240 = 1182$ proposed features, which are enumerated in more detail in Sections 3.2.1-3.2.5 of the text, respectively.
\end{table*}

\subsubsection{Basic Summary Features}
For each voice, we calculate five basic summary features from the \textit{Alicante} set: the number of notes, mean and standard deviation of the duration of all notes, and mean and standard deviation of the pitch of all notes \citep{de2007pattern}. Similarly to \citet{herlands2014machine}, we also calculate the proportion of notes and rests played simultaneously by all four voices. With the 5 monophonic features calculated for 4 voices plus the 2 polyphonic features, we have $5 \times 4 + 2 = 22$ basic summary features in total.

Such features may work together with the interval and sonata-inspired features to help reveal differences between Haydn and Mozart string quartets. The polyphonic features can indicate whether the voices interact differently in Mozart's versus Haydn's compositions. The interplay of voices is an important consideration in the string quartet genre, famously described by Johann Wolfgang von Goethe in 1829 as `a conversation among four intelligent people' \citep{klorman2016mozart}. The musicologist Charles Rosen (1997) also commented that `...perhaps the most striking innovation of Haydn's string-quartet writing' was its `air of conversation' (p. 141).

\subsubsection{Interval and Rhythm Features}
As previously mentioned, Mozart's music has been described as more `operatic' than Haydn's. According to Rosen (1997), `... Haydn brought all the weights of his symphonic experience to bear on his chamber music, as Mozart assimilated operatic and concerto style in his sonatas and quartets' (p. 45). While Mozart has a `breadth in dramatic writing' (Rosen, 1997, p. 80), Haydn exhibits a `keener sense of surface drama' than Mozart (Harutunian, 2005, p. 270). 
Interestingly, the `expressiveness' of Mozart has been discussed by biographers \citep{wallace}, economists \citep{borowiecki2017you}, and psychologists \citep{huguelet2005wolfgang};  Huguelet and Perroud even opined that Mozart's `worldwide recognition... [is] due not only to the genius of his music, but also to his ``temperament'' which has been portrayed in biographies and movies' (p. 130). Therefore, we propose intervallic features related to drama and expressiveness. We calculate the interval features discussed in this section (3.2.2) using pitch information on the full 1-132 scale, since octave information is necessary here (e.g., to distinguish rising and falling intervals).

\bigskip

\textbf{\textit{Pairwise Intervals.}}
We define \textit{pairwise} intervals by each pair of notes, in order.
For example, consider the excerpt of notes with pitches D, E, F, F, F, E, G, F. Then the corresponding pairwise intervals for pitch are D-E, E-F, F-F, F-F, F-E, E-G, G-F.  
These intervals are meant to identify local patterns, summarizing the relationships only between consecutive notes. Intervals defined by successive notes are included in \citep{de2007pattern} and often have been used for this task, e.g., (\citealp{kaliakatsos2011weighted,herlands2014machine,hontanilla2013modeling}).

For pitch tracks, we compute proportions of the following pairwise interval aspects for each voice of a movement: 
\begin{itemize}

\item The \textit{interval sign} specifies whether the interval is ascending, descending, or constant. For example, if the interval is middle C then the next E above, the interval would be labeled with ascending sign.
Interval signs are incorporated in the \textit{Jesser} feature set \citep{jesser1991interaktive}, among others.

\item The \textit{interval class} refers to the 12 possible distances in semitone (i.e., mod 12) on the chromatic scale: $0, 1, \hdots, 11$.
Summary statistics of interval class are frequently used as features, as in (\citealp{kaliakatsos2011weighted,herlands2014machine}).
We note that our usage of `interval class' is distinct from `interval class' in music theory, which only includes classes 0 to 6 due to inversional equivalence.

\item 
The interval's \textit{mode} refers to whether it is one of four categories: 
major, minor, perfect, or diminished/augmented. To calculate the mode, we map intervals with class $0, 1, \hdots, 11$ to P1, m2, M2, m3, M3, P4, d5/A4, P5, m6, M6, m7, M7, respectively. This mapping follows the most common representation of interval numbers and qualities in Western classical music, e.g., not all 3-semitone intervals are minor thirds, but augmented seconds are rare for music in the Classical period. While our interval mode features are mapped from the interval class features, they convey different information: the proportions are computed over different categories (4 for interval modes and 12 for interval classes), which allows for the possibility that, e.g., the proportion of major intervals generally distinguishes between the composers better than the proportion of major thirds specifically.
Summary statistics of nondiatonic intervals are included in the \textit{Alicante} feature set \citep{de2007pattern} and have been used in (\citealp{herlands2014machine, hillewaere2010string, taminau2010applying}). 
\end{itemize}
The proportions of 12 pitch classes, 3 signs, and 4 modes calculated for 
4 voices result in $(12 + 3 + 4) \times 4 = 76$ monophonic features here.

Drama and expressiveness are affected by the interval classes, signs, and modes, motivating their use for features. Interval sign has been linked to interval size. 
Large intervals create discontinuity in the melody, and ascending intervals heighten tension \citep{vos1989ascending}.
Therefore, large, ascending intervals are a frequent combination for drama, while small, descending intervals are combined for calm \citep{vos1989ascending}. Meanwhile, perception of happiness or sadness in music is related to mode \citep{temperley2013emotional}. Augmented and diminished modes can also contribute to dissonance, whose `most delicate implications' Haydn developed a `remarkable sensitivity' (Rosen, 1997, p. 130). In Haydn's and Mozart's string quartets, interval class and sign may reveal a difference in surface drama, while interval mode may expose a contrast in expressiveness.

We additionally compute the mean and standard deviation of pairwise interval semitones and differences in duration for each voice, as in \citep{de2007pattern}. These means and standard deviations contribute $2 \times 2 \times 4 = 16$ monophonic features.
For each pair of voices in a movement, the difference of those semitone means and the difference of those semitone standard deviations are calculated.  
In addition, voicepair differences in proportion of interval class are calculated. Voicepair differences are natural generalizations of monophonic features to polyphonic features and have been used in some studies, e.g., (\citealp{herlands2014machine, van2005musical}). These features, though simple, may reveal tendencies in Haydn's and Mozart's use of intervals and rhythm, particularly across voices. 
These differences of means, standard deviations, and 12 interval class proportions computed over all six possible 
pairs of voices add $(1 + 1 + 12) \times 6 = 84$ polyphonic features.

\bigskip

\textbf{\textit{Relative Pitch Intervals Within a Segment.}} \textit{Relative pitch} intervals are segment features defined by the first note of a segment and each subsequent note in the segment. For example, the segment from Figure \ref{fig:encoding} with length $m = 8$ notes D, E, F, F, F, E, G, F would have the relative pitch intervals D-E, D-F, D-F, D-F, D-E, D-G, D-F. Less local than the pairwise intervals, relative pitch intervals more effectively capture melodic context. To our knowledge, these intervals have never been used for the Haydn-Mozart string quartet classification task.

For a given segment length of $m$ notes (recall we consider 
$m=8,10, \ldots, 18$), relative pitch intervals are computed for each segment of pitches in the voice of a movement. Within each segment, we specifically calculate the proportion of minor third relative pitch intervals (i.e., defined as 3-semitone intervals, as above). In the example segment D, E, F, F, F, E, G, F, the calculated segment proportion of minor thirds would be 4/7.
The features are summary statistics of the segment proportions: minimum, first quartile, median, third quartile, maximum, mean, and standard deviation. Many segments contain no minor third intervals, while few segments contain mostly minor third intervals. Therefore, we include as features the count of segments with a low proportion (0) and a high proportion (at or above $0.6$). For each voice, $0.6$ is approximately the mean (over all movements) of the maximum proportion of minor third intervals. The 7 summary statistics of the proportions plus 2 count measures calculated for 4 voices and 6 segment lengths result in $(7 + 2)\times 4 \times 6 = 216$ relative pitch interval features. 

Minor third intervals are of special interest here for measuring expressiveness, because they contribute significantly to the perception of minor mode and a `sad' sound. Indeed, \citet{temperley2013emotional} found that listeners rate melodies containing a minor tonic triad (a type of chord containing a minor third) as sounding less happy than those containing a major tonic triad. The minor third is commonly used when modulating from a major key to a minor key.
The proposed minor third features can thus be associated with potential key modulations, providing a simple but automatic way to summarise quantitative evidence for whether Mozart's string quartets are more expressive than Haydn's.

\subsubsection{Exposition Features}
The exposition section of a sonata often contains an initial theme, the \textit{opening material}, followed by a secondary theme, the \textit{secondary material}. Occasionally, this convention is broken through \textit{monothematic} expositions. 
Harutunian (2005) and Rosen (1997) claim Haydn's sonatas are more often monothematic than Mozart's sonatas, and even a critic of Haydn's time `wrote admiringly that while less gifted composers needed many themes to sustain a movement, Haydn needed only one' (Rosen, 1997, p. 31).
Such claims motivate our proposition of exposition features related to repeated themes. 

We search for close repetitions of the opening material within the first half of each voice of a movement. This avoids detection of the recapitulation, which typically witnesses a repetition of the opening theme.
For a given segment length of $m$ notes, we compare the opening segment to all subsequent segments within the first half of a movement's voice. For all such pairs of segments, we compute the maximum fraction of overlap and its segment location (as defined in the beginning of Section 3.2).
If there are multiple segments with the same maximum fraction, then the location is defined using the later segment in the piece, as a way to measure the extent to which the opening material is sustained through the exposition. The fraction of overlap count is computed for thresholds $t = 0.7,0.9$, and $1$. 
Besides exact matches (i.e., with threshold 1), segments with a high degree of similarity (i.e., with thresholds at or above $0.7$ or $0.9$) are of interest, since listeners would likely perceive the segments as sounding approximately the same. The exposition features are calculated separately for notes' pitches and duration. With the maximum fraction of overlap, its location, and 3 fraction of overlap counts calculated for 4 voices, 6 segment lengths, and both pitch and duration tracks, there are $(1 + 1 + 3) \times 4 \times 6 \times 2 = 240$ exposition features.

If Haydn is more likely than Mozart to have monothematic expositions, then we might expect his sonatas to yield higher maximum fractions of overlap, locations, and threshold counts than Mozart. A fraction of overlap equal to $1$ indicates a perfect repetition of the opening material within the exposition, so a high count at threshold $1$ suggests one recurring theme. A location with a high value may reflect a theme sustained throughout the exposition, potentially corresponding to monothematicism.

\subsubsection{Development Features}
The exposition section of a sonata leads into the development section, which contains exploration and contrast of the beginning themes. Haydn and Mozart may differ in their development styles: Harutunian (2005) asserts that Mozart exhibits more `continuous flow' from the exposition into the development, while Haydn possesses `an immediate formal delineation' between the two sections (p. 199). To identify such differences, we propose features related to musical variability: these features are therefore distinct from the exposition, where the focus is repeated themes.

For each segment of length $m$ notes within the voice of a movement, we compute the standard deviation of notes within the segment. The maximum of all such standard deviations and its segment location are calculated. If multiple segments have the same maximum standard deviation, the location is determined by the first occurrence, so that the start of greatest musical contrast can be roughly identified within the development (recall that these computations are fully automated, calculated simply from the symbolic representations of scores). We also measure the extent of variability within the voice of a movement using counts of segments with large standard deviations, i.e., exceeding various thresholds $s$. Appropriate values for $s$ are chosen as weighted $0.70,0.80,0.90$, and $0.95$ quantiles of the standard deviations $\textrm{sd}(\cdot)$, with the weights defined to account for differing movement lengths. More formally, for each segment length $m$, voice $j$, and quantile $q$, we compute the threshold $s_{jm}(q)$ as the $q$-th quantile of the set $\{\frac{1}{M_i - m + 1} \textrm{ sd(notes in segment$_{ijkm}$)}: k = 1, 2, \hdots, M_i - m + 1, i = 1, \hdots, n \}$, where $M_i$ is the number of notes in movement $i$, so that $M_i - m + 1$ is the number of segments. The development features are calculated separately for notes' pitches and durations.
With the maximum standard deviation, its location, and 4 count measures for 4 voices, 6 segment lengths, and both pitch and duration tracks, there are $(1 + 1 + 4) \times 4 \times 6 \times 2 = 288$ development features in total.

If Haydn's developments consist of more `organic construction' and `greater sectionalization' \cite[p. 273-4]{harutunian2005haydn}, then these aspects may translate to, on average, Haydn string quartets having higher maximum standard deviations and counts. The location of a maximum standard deviation could map the start of the greatest change within a movement's development; differences between Haydn's and Mozart's locations may suggest distinct placements of peak tumultuous material.

\subsubsection{Recapitulation Features} 
In the recapitulation, the material from the exposition is often reiterated. Harutunian (2005) claims, `Mozart's recapitulations mirror his expositions far more closely than do Haydn's' (p. 212); his changes are often `ornamental', unlike Haydn's `sweeping changes' (p. 270). Therefore, we measure aspects related to the recapitulation and how closely it would match the exposition. These are similar to the exposition features, but applied to locate the final repetitions of opening material.

For a given segment length of $m$ notes, we compare the opening segment to all subsequent segments in the voice of a movement. For each segment, we calculate the fraction of overlap. The maximum fraction of overlap over the segments and its location become our features. In the case of multiple segments with the same maximal fraction, the location is determined by the final occurrence, as a way to capture the symmetry between the end of a piece and its opening material. The fraction of overlap counts at thresholds $t = 0.7,0.9$, and $1$ are computed. 
Our rationale for choosing these thresholds is similar to that for the exposition thresholds. Like the other sonata-inspired features, the recapitulation features are calculated separately for notes' pitches and durations. With the maximum fraction of overlap, its location, and 3 fraction of overlap counts calculated for 4 voices, 6 segment lengths, and both pitch and duration tracks, there are $(1 + 1 + 3)\times 4 \times 6 \times 2 = 240$ recapitulation features. 

The maximum fraction of overlap and counts can potentially measure similarity between the exposition and recapitulation sections. Higher values for these features in Mozart compositions, on average, could verify Mozart's exposition-recapitulation symmetry. The location of the maximum fraction of overlap is the last closest repetition of opening material within the voice of the movement; as such, it may indicate differences in Haydn's versus Mozart's approach to concluding a piece. 

\section{Statistical and Machine Learning Methods}
Using the musical features from the previous section, we apply interpretable statistical and machine learning methods to analyse the differences between Haydn and Mozart string quartets. In 4.1, we propose our classification model. In 4.2, we discuss feature selection. The interested reader may refer to Hastie, Tibshirani, and Friedman (2001) for an excellent overview of methods and issues in statistical machine learning.

\subsection{Classification Model}

For the classification model, we use logistic regression, which is perhaps the most well-known model for binary response data \citep{agresti2003categorical}. Logistic regression is a type of generalized linear model that can estimate the probability of an observation belonging to one of two classes, given the values of a set of features. Advantages of this model include its ease of interpretation (i.e., the effect of each feature on the composer probability can be clearly explained) and the availability of well-understood inference procedures. We assume the usual additive effects, so that mathematically the model is of the form 
\begin{equation}
\label{eq:glm}
\pi(X)=\frac{e^{\beta_0 + X\beta}}{1+e^{\beta_0 + X\beta}},
\end{equation}
where $\pi$ is the probability of a movement belonging to the Haydn versus Mozart class, $X$ is the $n \times p$ data matrix containing the $n$ movements and $p$ features, $\beta_0$ is the intercept, and $\beta = (\beta_1, \beta_2, \ldots, \beta_p)'$ is a $p \times 1$ vector of coefficients for the features. Then $\beta_0$ and $\beta$ are the model parameters to be estimated from training data, which may be interpreted as follows: the intercept $\beta_0$ is associated with the probability $e^{\beta_0} / (1+e^{\beta_0})$ that the movement is by Haydn if all the feature values are zero, and the coefficients of $\beta$ indicate the contribution of each feature to the classification of a movement as by Haydn (when the coefficient is positive) or Mozart (when the coefficient is negative).

For improved numerical stability in parameter estimation, Bayesian logistic regression is used from \citep{gelman2008weakly}. To each coefficient except the intercept, independent Cauchy prior distributions with mean $0$ and scale $\xi / (2\hat{\sigma})$ (where $\hat{\sigma}$ is the sample standard deviation of the associated feature and $\xi$ is the prior scale factor) are applied; for the intercept, a more conservative Cauchy prior distribution with mean $0$ and scale $10$ is used. Implementation is provided through the \textit{bayesglm} function from the R package \textit{arm} \citep{gelman2009arm}, where the default prior scale factor is $\xi=2.5$.  In practice, setting $\xi$ is one approach to control the number of features to be included in the final model after feature selection (Section 4.2), known as \textit{regularization} in the statistical literature, e.g., as reviewed in \citep{polson2012shrink}. The role of $\xi$ in our methods is further discussed below in Section 4.2.

To use a logistic regression model for classification, the estimated probabilities $\hat{\pi}(X)$ must be converted to binary classes. In datasets with balanced classes, it is customary to apply a cutoff of 0.5, which we follow for HM107: assign observations with greater than $0.5$ (estimated) probability to one class, and the remaining observations to the other class.
In a dataset with imbalanced classes as in HM285, a cutoff $0.5$ may not be optimal for classification accuracy; instead the cutoff may be treated as a tuning parameter, which is a type of approach that has been   explored for such binary classification problems \citep{zou2016finding}. We test the sequence of cutoff values $0, 0.01,0.02,\hdots,0.98,0.99,1$ and choose the `best' cutoff value as the one that maximises classification accuracy within the training data.

To pre-process the data, we remove computed features that have variability $\approx$ 0, as is often done prior to feature selection and classification \citep{kuhn2008building}. From the 1182 proposed features, this results in 1115 features for HM285 and 1110 features for HM107, on which feature selection will be applied.

\subsection{Feature Selection}

The goal of feature selection is to determine the appropriate features to include in the logistic regression model for classification. From a practical perspective, feature selection helps identify a succinct subset of features representing meaningful differences between Haydn and Mozart. From a statistical perspective, feature selection is essential here, because we have over 1000 features and sample sizes of only 107 or 285 movements; estimation of unique model coefficients in Equation \ref{eq:glm} at minimum requires the sample size to exceed the dimension (i.e., number of features), and a common guideline for generalized linear models is that the sample size should be at least ten times the dimension \citep{agresti2003categorical}. Further, a fitted logistic regression model that contains a large number of features could be difficult to interpret and suffer from reduced classification accuracy due to overfitting. In particular, highly collinear features included together in a fitted model would not have meaningful interpretations for their estimated coefficients, due to inflated standard errors.
For example, the `minor third relative pitch interval' features include summary statistics such as the mean and median, which are strongly correlated; if these minor third features are important for classification, then feature selection can identify the optimal one to include in the model.

There are many feature selection approaches from the statistical and machine learning literature, including methods that transform the features to reduce their dimensionality (e.g., factor analysis, principal component analysis, and discriminant analysis) and subset selection algorithms that search for optimal subsets of features (e.g., stepwise regression) \citep{guyon2003introduction}. Our proposed features have musical meaning that would be lost in a transformation, so we use subset selection. Specifically, we aim to identify which subset of the $p$ features should be included 
to yield the ‘best’ logistic regression model in Equation \ref{eq:glm}. We opt for the Bayesian information criterion (BIC) \citep{schwarz1978estimating} to judge the quality of a subset of features, as the BIC is a standard criterion used for model selection in statistics. From a fitted logistic regression model, the BIC is computed as
\begin{equation}
\textrm{BIC}=-2\mathcal{L}+2 (d+1) \log (n),
\end{equation}
where $n$ is the number of observations in the dataset, $d \leq p$ is the number of features included in the model, and $\mathcal{L}$ is the maximised value of the log-likelihood of the model fitted with those features. The subset of features that leads to the lowest BIC value in the fitted model would be considered the `best' subset of features.

However, it is not computationally feasible to exhaustively test all possible subsets of features to find the one with the lowest BIC; we note there are $2^{p}$ such combinations. In practice then, one can only test a limited number of subsets and choose the model with the lowest BIC value found. We use the method of Iterative Conditional Minimization (ICM) to search for the minimum BIC, which is discussed in \citep{zhang2007lookahead} as a simple but substantively more effective alternative to stepwise regression methods for subset selection. In addition to the model selection criterion (which we have chosen as BIC), ICM requires a choice of the initial ordering of features: ascending order, descending order, or random order. To reduce the risk of ICM identifying a poor local solution, we choose random ordering, run the algorithm for 10 different random orderings, and select the lowest BIC model from among the 10 resulting models. The algorithm is summarised in pseudocode in Algorithm 1.

\begin{tcolorbox}[title = {Algorithm 1. Feature Selection}]
\textbf{For} $k = 1, \hdots, 10$:
\begin{itemize}
   \item[]
    \textbf{Initialise:}
    \begin{itemize}
	    \item[] Set  $S_k$ to be an empty set and BIC$_k = +\infty$.
	    \item[] Randomly order the $p$ features $x_1, x_2, \hdots, x_p$ from the full feature set as $x_{(1)}^{k}, x_{(2)}^{k}, \hdots, x_{(p)}^k$.
    \end{itemize}
 
   \textbf{For} $j = 1,2, \ldots, p$:
    \begin{itemize} 
        \item[] \textbf{If} $x_{(j)}^k$ is not in $S_k$, \textbf{then} 
        \begin{itemize}
            \item[] Fit a logistic regression model with 
            $x_{(j)}^k$ and all features from $S_k$.
            \item[] \textbf{If} the BIC from this fitted model is less than BIC$_k$, \textbf{then} add $x_{(j)}^k$ to $S_k$. 
        \end{itemize}
       \item[] \textbf{Else if} $x_{(j)}^k$ is in $S_k$, \textbf{then} 
            \begin{itemize}
                \item[] Fit a logistic regression model with 
                all features from $S_k$, excluding $x_{(j)}^k$.
                \item[] \textbf{If} the BIC from this fitted model is less than BIC$_k$, \textbf{then} remove $x_{(j)}^k$ from $S_k$.
            \end{itemize}
        \end{itemize}
    Repeat \textbf{For} loop until two successive passes yield no further additions or deletions of features to $S_k$.
    \newline
    \textbf{Return} $S_k$ and $BIC_k$.
\end{itemize}
Let $S = S_b$, where $b = \underset{k = 1, 2, \hdots, 10}{\textrm{argmin}}\textrm{BIC}_k$.
\newline
\textbf{Return} $S$, the reduced feature set to be used for classification.  
\end{tcolorbox}

The number of features selected in the subset by Algorithm 1 will depend on the prior scale factor $\xi$ used for fitting the Bayesian logistic regression model: larger values of $\xi$ lead to less regularization and more features being selected. The default of $\xi=2.5$ in \textit{bayesglm} yields models with 22 to 26 features on the HM285 dataset. We selected $\xi=0.6$ for our analysis, so that the number of features selected for both datasets is consistently fewer than 10, in accordance with the guideline for model dimension in \citep{agresti2003categorical}.

\section{Results and Discussion}

In 5.1, we present our cross-validated composer classification results from applying the statistical machine learning methods from Section 4 on the HM107 and HM285 datasets; our predictive results are then compared to prior studies. The remaining sections consist of additional analyses on the larger dataset, HM285. In 5.2, we perform analyses of statistical stability.
In 5.3, we summarise one comprehensive model of composer. In 5.4, we discuss the musical interpretations gained from our model.

\subsection{Classification Results with Comparison to Prior Studies}
Following the statistical approach outlined in the previous section, we classify the composer of Haydn and Mozart string quartets in the HM107 and HM285 datasets. The resulting classification accuracies can serve as a `reality check' before interpreting the models \citep[p. 3920]{yu2020veridical} and can quantify the extent to which these compositions can be discriminated.

As highlighted in \citep{hillewaere2010string}, the datasets are small, motivating the use of cross-validation (CV) to assess classification ability. We choose the leave-one-out (LOO) CV approach, which estimates the `true' classification accuracy for unseen observations using observations from the dataset \citep{arlot2010survey}. In past studies, LOO was the most common CV approach for the Haydn-Mozart string quartet task, so our use of LOO facilitates comparison. The evaluation criterion is the classification accuracy.

LOO is described as follows. For each movement $i=1,2,\hdots,n$, we form training fold $i$ and testing fold $i$. Training fold $i$ contains all movements except $i$, and testing fold $i$ contains only movement $i$. Feature selection and model fitting are performed on training fold $i$, and the composer is subsequently predicted for movement $i$. That is, within training fold $i$, the `best' subset of features is identified from feature selection using Algorithm 1, a Bayesian logistic regression model is fit with that subset of features, and a classification cutoff is selected (i.e., 0.5 for HM107 and tuned to maximise training classification accuracy for HM285); subsequently, the same fitted model and cutoff are used to classify the composer of movement $i$ in testing fold $i$. The LOO classification accuracy is computed here as the proportion of movements correctly classified in their respective testing folds. 
Hence, LOO involves fitting and predicting from $n$ models, one per fold.

\begin{figure}
\includegraphics[width=1\textwidth,trim={0cm .1cm 0cm 0cm},clip]{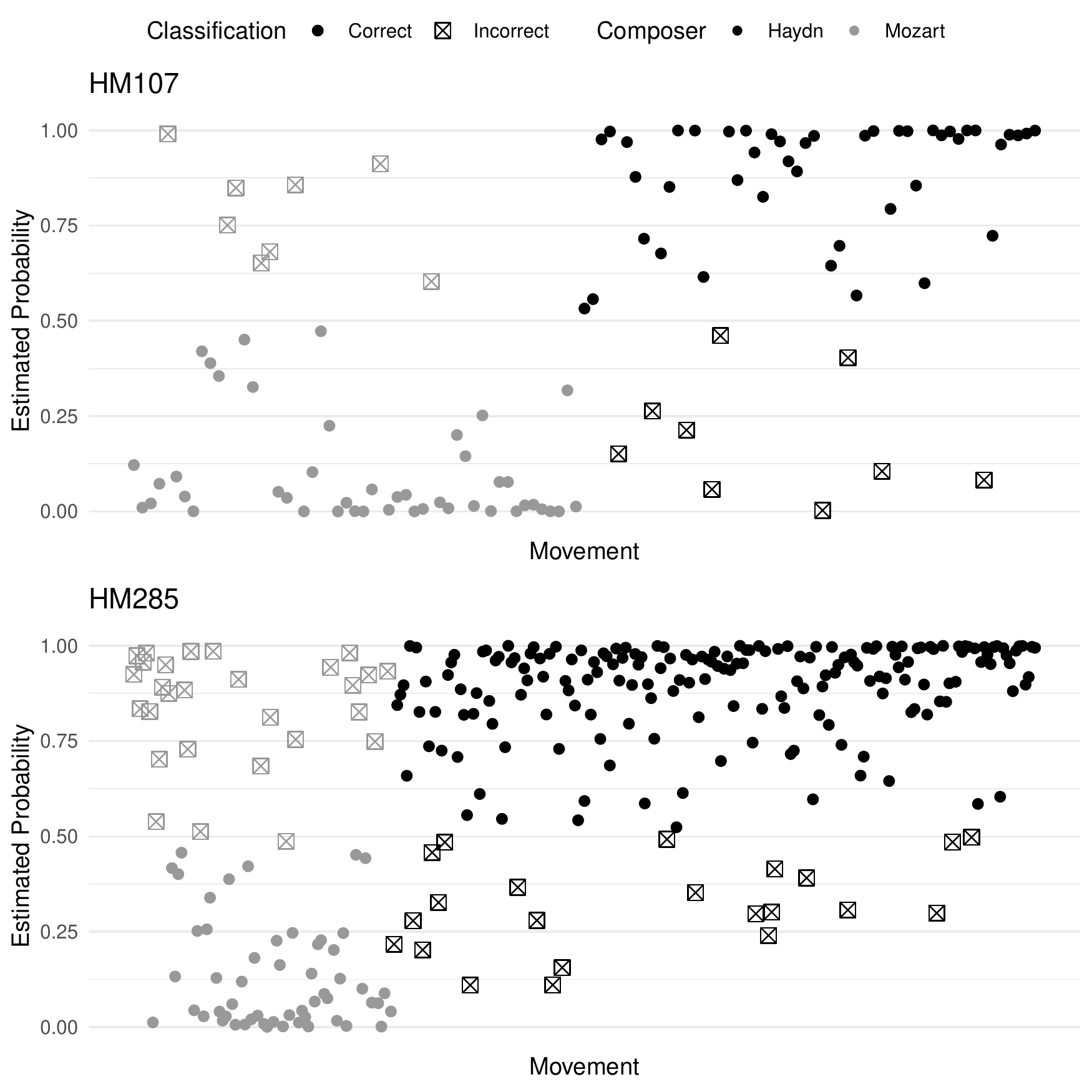}
\centering
\caption{For the HM107 (top) and HM285 (bottom) datasets, we plot the estimated composer probability from logistic regression for each left-out movement in LOO CV. Movements (on the horizontal axis) have been ordered by approximate date of composition for each composer. Symbols mark whether the movement was classified correctly or incorrectly, and colour indicates the true composer of the movement. Both datasets exhibit strong separation in estimated composer probability. In HM107, the mean probability for Mozart movements (class label 0) is close to 0 at about 0.2043, while the mean probability for Haydn movements (class label 1) is near 1 at approximately 0.7708. In HM285, the mean probability for Mozart movements is 0.3740, which is much lower than the mean probability 0.8322 for Haydn movements.
}
\label{fig:jitter}
\end{figure}

For LOO on the larger dataset HM285, our approach achieves a classification accuracy of approximately 82.46\%. The LOO Haydn accuracy, which we define here as the proportion of Haydn movements correctly classified in LOO, is very high at about 89.16\%. The LOO Mozart accuracy, defined in an analogous manner, is about 65.85\%. A lower accuracy rate for the Mozart class is not surprising, since there are more than twice as many Haydn as Mozart movements in HM285. Still, as summarised in Table \ref{table:confusion_matrix}, relatively few movements are misclassified overall: 28 Mozart and 22 Haydn. 
The fitted probabilities from the LOO testing folds are plotted in Figure \ref{fig:jitter} (bottom panel).
Generally, the composers are well-separated, with many probabilities clustering around the correct composer (0 for Mozart, 1 for Haydn).

For LOO on the smaller dataset HM107, the LOO classification accuracy is about 84.11\%, higher than attained in prior studies using the same dataset (Table \ref{table:lit_acc}). The predictive performance is strong for both composers, since the LOO Haydn and Mozart accuracies reach approximately 83.33\% and 84.91\%, respectively. Only 17 movements in total are misclassified in this dataset: 8 Mozart movements and 9 Haydn movements (Table \ref{table:confusion_matrix}). As in HM285, the fitted probabilities in HM107 exhibit strong separation by composer (Figure \ref{fig:jitter}, top panel).
Altogether, the well-separated probabilities and high accuracies on the HM107 and HM285 datasets suggest differences between Haydn and Mozart string quartets that are detected by our statistical machine learning methods.

\begin{table}[h]
	\centering
    \caption{Confusion Matrix for Haydn-Mozart String Quartet Classification with Bayesian Logistic Regression and LOO CV}
    \label{table:confusion_matrix}
	\begin{tabular}{lllll}
    \toprule
Dataset & &Predicted Mozart & Predicted Haydn\\
    \midrule
HM107   &Observed Mozart &45 &8\\
        &Observed Haydn &9 &45\\ 
    \midrule
HM285   &Observed Mozart &54 &28\\
        &Observed Haydn &22 &181\\ 
\bottomrule
\end{tabular}
\end{table}

We compare our results and methods to all existing studies of which we are aware at the time of writing in May 2019 (Table \ref{table:lit_acc}).
In the pioneering work, the authors implement a 3-nearest neighbour classifier with 20 lower-level musical features transformed through Fisher's discriminant analysis \citep{van2005musical}. Next, \citet{hillewaere2010string}, \citet{hontanilla2013modeling}, and \citet{kaliakatsos2011weighted} use $n$-gram (or $(n-1)$th order Markov models) from language analysis, modeling the probability of a musical event given the context of past musical events. \citet{herlands2014machine} classify composer with either linear SVM or the Naive Bayes classifier. \citet{velarde2016composer} attain the highest accuracy on HM107 prior to our study, but they use a computer vision approach that is difficult to interpret musically: they apply a Gaussian filter to images of piano roll scores, transform the resulting pixel data through linear discriminant analysis, and classify with a linear SVM. In follow-up work, \citet{velarde2018convolution} extend their approach to include image analysis of spectrograms, as well as classification with a $k$-nearest neighbour classifier; however, as before, their study differs in scope from our interpretable machine learning study.
Finally, \citet{taminau2010applying} deploy subgroup discovery, a descriptive rule learning technique that involves both predictive and descriptive induction. Evidently, a diverse range of approaches have been applied to the Haydn-Mozart classification problem. 
It is interesting that all the reported classification accuracies in Table \ref{table:lit_acc} exceed the aforementioned human accuracy of $\approx$ 67\%, with many close to 80\%. These results raise potential research questions regarding the `inherent difficulty' of the Haydn-Mozart string quartet classification problem and other tasks in MIR, i.e., is a given problem `fundamentally intractable' for algorithms to learn beyond a certain accuracy? \citep[p. 256]{jordan2015machine}.

\begin{table}[h]
    \small
    \renewcommand{\tabcolsep}{3pt}
    \caption{Comparing Classification Accuracy Rates and Methods with Prior Studies}
    \label{table:lit_acc}
	\begin{tabular}{llll}
    \toprule
Method &Dataset &Cross-validation &Accuracy \\
\midrule
FDA + k-means clustering (Van Kran. et al., 2005) &HM107 &LOO &79.44\% \\
LDA + Linear SVM \citep{velarde2016composer} &HM107 &LOO &80.4\% \\
KNN + SVM ensemble (Velarde et al., 2018) &HM107 &LOO &74.8\%\\
\textbf{ICM + Bayesian Logistic Regression (ours)} &HM107 &LOO &84.11\%\\
\midrule 
3-grams model (only Cello) \citep{hillewaere2010string} &*
&LOO &75.4\% \\
Weighted-Markov chain model + SVM (Kal. et al., 2011) &* &30 simulations &70\% \\
Linear SVM or Naive Bayes \citep{herlands2014machine} &* &CV trials &80\% \\
3-grams model \citep{hontanilla2013modeling} &* &LOO &74.7\%\\
Subgroup discovery \citep{taminau2010applying} &*
&LOO  &73\%\\
\midrule
\textbf{ICM + Bayesian Logistic Regression (ours)} &HM285 &LOO &82.46\%\\
\bottomrule
\end{tabular}
An asterisk in the Dataset column indicates that a study used a different dataset than HM285 or HM107.
\end{table}

\subsection{Statistical Stability of Classification Results}
Having established the strong predictive performance of our approach according to the traditional LOO metric, we perform additional analyses on the larger dataset, HM285. In this section, we explore our approach's statistical stability (i.e., consistency of results after reasonable perturbations, such as to the data through CV), which can be viewed as a `a prerequisite for
trustworthy interpretations' \citep[p. 2]{murdoch2019interpretable}.

\begin{figure}
\includegraphics[width=1\textwidth]{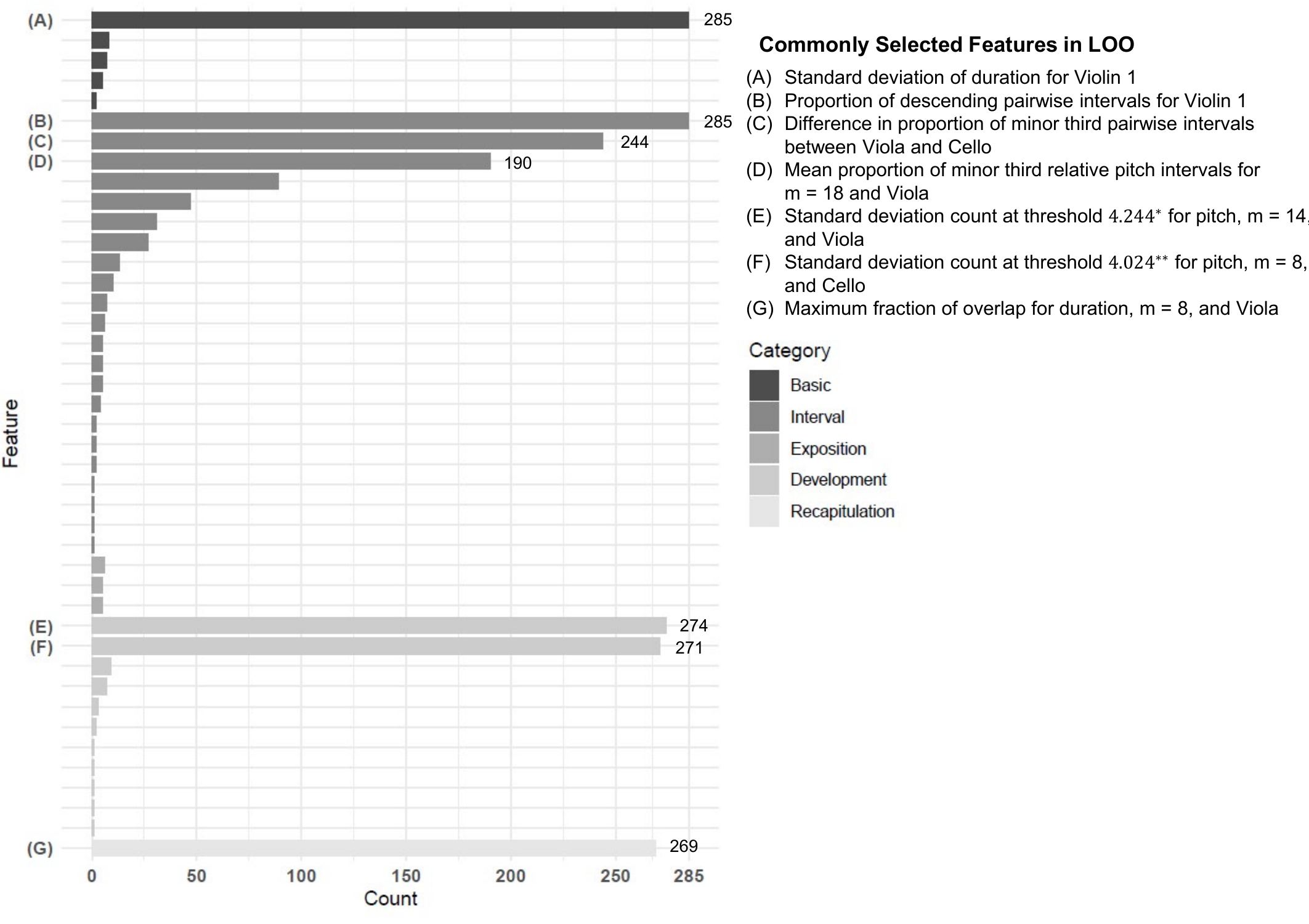}
\centering
\caption{The barplot shows the number of times a feature is selected in LOO CV in the HM285 dataset, from among the 42 features selected in at least one fold. Colours correspond to feature category and demonstrate that all feature categories are represented in LOO CV. The features labeled (A)-(G) are the most commonly selected features, i.e., identified in more than 190 folds. For the development features (E) and (F), the thresholds are the following weighted quantiles: $^* 0.80$ and $^{**} 0.70$.}
\label{fig:LOO_barplot}
\end{figure}

First, we examine the stability of selected features across the 285 LOO folds in HM285. 
The subsets of features selected from random ICM (Algorithm 1) are very similar across the folds. In total, there are 42 features selected in at least one fold of LOO, summarised in Figure \ref{fig:LOO_barplot}. As listed in the figure, seven of those features are included in 190 or more folds, with two included in all 285 folds.
These commonly selected features confirm the stability of our feature selection approach across LOO folds and suggest potentially important features for interpretation.

Second, we confirm the stability and validity of cross-validation, which assumes that for each pair of training and testing folds, observations across folds are independent and identically distributed according to some probability distribution \citep{arlot2010survey}. The dataset contains natural groupings that could violate the assumption of independence, particularly that each movement is one of several in a quartet.
Thus, we explore an alternative CV scheme, namely \textit{leave-one-quartet-out} (LOQO), which can help assess the stability of our classification results after accounting for within-quartet dependence.
The LOQO CV scheme has not been considered in prior Mozart-Haydn MIR studies. Instead of leaving out one movement in each testing fold as in LOO, LOQO leaves out one quartet in each testing fold. For example, Mozart's No. 4 string quartet (K. 157) has three movements, which would all be left out in a LOQO testing fold. 
The LOQO classification accuracy is computed as the mean classification accuracy over the 86 testing folds, i.e., the 86 left-out quartets in the HM285 dataset. The resulting LOQO classification accuracy is about 79.03\%. Details by movement can be found in the Supplementary Material, which shows that
the estimated probabilities, classes, and selected features from LOQO and LOO are often very similar. In particular, the most commonly selected features in LOQO are the same seven features listed in Figure \ref{fig:LOO_barplot}. Additionally, we conducted descriptive analyses to explore other groupings in the dataset but did not find evidence of dependence among movements by set, quartet, or date of composition (included in the Supplementary Material). We conclude that the LOO and LOQO CV schemes do not violate the independence assumption and give generally stable results.

Finally, we repeat the LOO CV scheme on a reduced subset of the features, which contains only the basic summary and interval features. By excluding the novel sonata-inspired features (exposition, development, and recapitulation), we can better understand their role in prediction. The resulting LOO classification accuracy is 80.35\%, while the LOO Haydn and Mozart accuracy rates are about 90.15\% and 56.10\%, respectively. This represents 2 out of 203 more Haydn movements classified correctly than before and 8 out of 82 more Mozart movements classified incorrectly, meaning that more movements in the HM285 dataset are guessed to be the majority class, Haydn. Overall, LOO on the reduced subset results in close to a 2\% decrease in LOO classification accuracy. Comparing the movements misclassified from LOO on the full feature set versus on the reduced feature set can offer insights into potentially important features and the contribution of structural information.
As discussed in the Supplementary Material, this analysis suggests that the `maximum fraction of overlap' recapitulation feature and the `standard deviation count at threshold 4.244' development feature (the same two features listed in Figure \ref{fig:LOO_barplot}) are particularly influential sonata-inspired features for improved classification. It is interesting that feature selection did not commonly identify exposition features, considering several musicologists have commented that Haydn's expositions are more often monothematic than Mozart's (\citealp{harutunian2005haydn, rosen1997classical}); potential reasons for this are discussed in Section 5.4.

\subsection{Model of Composer on Musical Features}
Having confirmed the strong predictive performance and assessed the statistical stability of our methods, we fit a single model of composer for interpretation. The statistical machine learning methods of Section 4 are applied to all 285 movements of the HM285 dataset together, and we summarise the resulting model in what follows. 

For each feature $j$ (along with the intercept) in the model, the estimated coefficient $\hat{\beta}_j$ and its standard error $\sqrt{\hat{Var}(\hat{\beta}_j)}$ are given in Table \ref{table:full_model}. Each feature's coefficient corresponds to a change in probability of composer and a contribution to classification, as previously shown in Equation \ref{eq:glm} and discussed in Section 4.1. 
For coefficients with positive sign, increases in the feature correspond to a greater probability the movement is composed by Haydn, adjusting for other model features. For example, $\hat{\beta}_4=44.03$, so Haydn is more likely than Mozart to have higher mean proportions of minor third intervals in the viola voice, controlling for the other features. In contrast, we interpret features with negative coefficients as negatively associated with Haydn. For example, Haydn movements are less likely than Mozart movements to have high standard deviations of duration values in the Violin 1 voice (since $\hat{\beta}_1 = -15.47$), adjusting for other features.

By the assumption of additivity, a coefficient's effect on composer is constant for each value of the feature, even as other features' values change. The effect of each feature (including, trivially, the intercept) on composer is tested by the hypotheses
\begin{equation}
\begin{aligned}
H_0&: \beta_j =0 \textrm{ when all other features are in the model}\\
H_A&: \beta_j \neq 0 \textrm{ when all other features are in the model},
\end{aligned}
\end{equation}
for $j=0, 1,\hdots,7$.
The Wald $p$-values for these tests are listed in Table \ref{table:full_model}. For each coefficient except the intercept, the $p$-value is less than $0.01$. 
Strongly significant $p$-values are a natural consequence of the use of BIC as the model selection criterion. For example, the `standard deviation counts at thresholds 4.244 and 4.024' have $p$-values below $10^{-8}$, indicating these counts have significant effects on composer. 

Most commonly, a logistic regression model's goodness of fit is assessed through deviance, a generalization of analysis of variance \citep{nelder2004generalized}. Here, the deviance would compare the maximised log-likelihood for the fitted model and for the saturated model (which contains as many parameters as observations). This is handled in our case by using BIC for feature selection, since BIC is a function of the maximised log-likelihood of the fitted model.
Tests based on residuals can also be used, and here we apply the Hosmer-Lemeshow test.

In the Hosmer-Lemeshow test \citep{hosmer1980goodness}, the estimated probabilities from the model are divided into $g$ groups, in which the observed outcomes are compared to the expected outcomes from the model. When the model fits the data well and $g$ is chosen such that $g > p+1$, the test statistic has an approximate $\chi^2$ distribution. We test values of $g$ ranging from 20 to 100. All tests yield $p$-values greater than 0.5, and the median $p$-value is 0.9880. With generally large $p$-values over the range of $g$, there is no evidence of lack of fit.

\begin{table}[h]
    \scriptsize
    \caption{Additive Bayesian Logistic Regression Model of Composer on Musical Features for HM285 Dataset}
    \label{table:full_model}
    \resizebox{\linewidth}{!}{
	\begin{tabular}{lllll}
    \toprule
Category &Feature &$\hat{\beta}_j$ &$\sqrt{\hat{Var}(\hat{\beta}_j)}$&$p$-value\\
\midrule
&(Intercept) &-1.12 &2.32 &0.63\\
Basic summary &Standard deviation of duration for Violin 1 &-15.47 &4.23 &0.00025\\
Interval &Proportion of descending pairwise intervals for Violin 1 &16.71 &3.85 &$1.44 \times 10^{-5}$\\
Interval &Difference in proportion of minor third pairwise intervals between Viola and Cello &-16.53 &4.35 &0.00014 \\
Interval &Mean proportion of minor third relative pitch intervals for $m=18$ and Viola &44.03 &10.98 &$6.03 \times 10^{-5}$\\
Development &Standard deviation count at threshold $4.244^{*}$ for pitch, $m=14$, and Viola &0.094 &0.015 &$1.15 \times 10^{-9}$\\
Development &Standard deviation count at threshold $4.024^{**}$ for pitch, $m=8$, and Cello &-0.027 &0.0046 &$3.80 \times 10^{-9}$\\
Recapitulation &Maximum fraction of overlap for duration, $m=8$, and Viola &-4.66 &1.78 &0.0087\\  
\bottomrule
\end{tabular}
}
\scriptsize  
The intercept $\hat{\beta}_0$ is associated with the probability that the movement is by Haydn, if all feature values are zero. For $j \neq 0$, each coefficient $\hat{\beta}_j$ indicates the contribution of a feature to the classification of a movement as by Haydn (when $\hat{\beta}_j > 0$) or Mozart (when $\hat{\beta}_j < 0$). The estimated standard error $\sqrt{\hat{Var}(\hat{\beta}_j)}$ is used in calculating the $p$-value to assess the statistical significance of a coefficient. 
\newline 
Note that for the development features, the thresholds are the following weighted quantiles: $^*0.80$ and $^{**}0.70$.
\end{table}

\subsection{Musical Interpretation}
\begin{figure}
\includegraphics[width=.999999\textwidth]{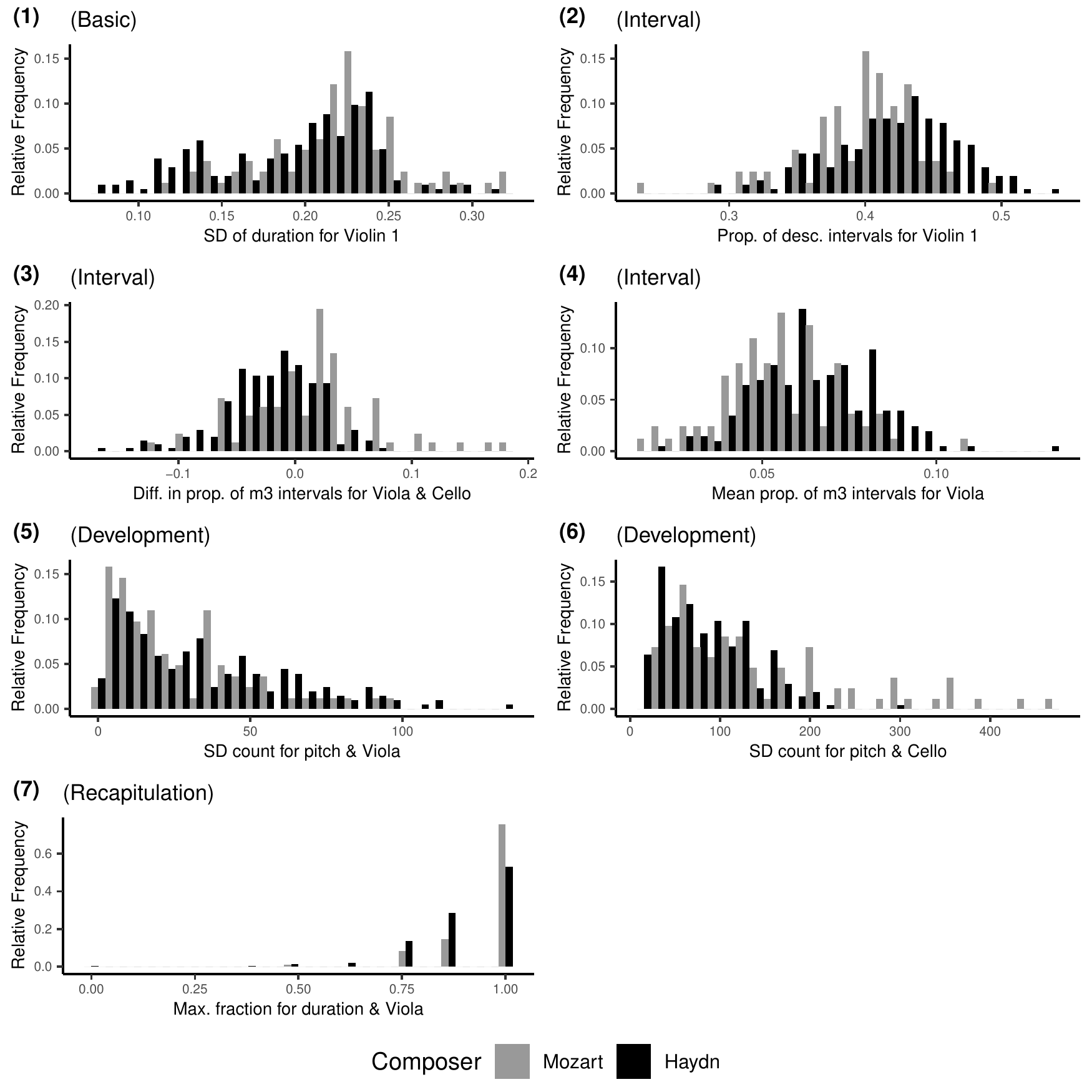}
\centering
\caption{ 
Side-by-side relative frequency plots by composer summarise
seven features from the full model, each of which was identified from feature selection as a `good' discriminator of composer on the HM285 dataset. Each feature's plot corresponds to a row of Table \ref{table:full_model}, e.g., the `standard deviation of duration for Violin 1' feature in plot (1) has coefficient $\hat{\beta_1} = -15.47$ in the table.
}
\label{fig:rel_freq}
\end{figure}

We now provide musical interpretations of the differences between Haydn and Mozart string quartets, based on the features from the model in the previous subsection. That model was identified from feature selection as a `good' discriminator of composer in the HM285 dataset, suggesting differences in the features for Haydn versus Mozart. The features were also frequently selected in LOO CV in Section 5.1 and stability analyses in Section 5.2, further motivating their analysis. 
The distribution of each feature is plotted by composer in Figure \ref{fig:rel_freq}. Our results and interpretations for the features generally agree with the musicological claims discussed in Section 3 regarding Haydn-Mozart compositions. We discuss these features in detail, starting with basic summary and interval features, followed by sonata-inspired features.

For the basic summary features, the standard deviation of duration for Violin 1 tends to be higher for Mozart than for Haydn (Figure \ref{fig:rel_freq} (panel 1)). For intervals, several differences between these composers are identified. First, higher proportions of descending pairwise intervals in the first violin are associated with Haydn, rather than Mozart (panel 2), potentially confirming differences in `surface drama'.
Meanwhile, the difference in proportion of pairwise intervals with minor third interval class between Viola and Cello is generally higher for Mozart than Haydn (panel 3), implying that Mozart's handling of minor third intervals is more distinctive between the low accompanying voices than Haydn's. 
For the relative pitch intervals,
Haydn tends to have a higher mean proportion of minor thirds than Mozart in Viola (panel 4). The inclusion of two minor third interval features suggest possible distinctions in `expressiveness' or `drama' between Haydn and Mozart, as introduced in Section 3.2.2.

The three sonata-inspired features in the model point to differences between the composers in some areas highlighted by musicologists.  
In Figure \ref{fig:rel_freq} (panel 5), Haydn's greater standard deviation counts at a high threshold and segment length (the 0.80 quantile and $m = 14$ notes) in the viola voice could correspond to his `organic construction' in the development.
Meanwhile, for a different voice (Cello) and a lower threshold and segment length (the .70 quantile and $m = 8$ notes) in panel 6, Mozart generally has higher standard deviation counts than Haydn. Thus, these two features align with musicological claims that Haydn and Mozart could differ in the extent of thematic material variation in the development, but the direction of the effect is inconclusive.
The recapitulation maximum fraction of overlap for duration in Viola has a negative coefficient (Table \ref{table:full_model}) and is more often 1 for Mozart than Haydn (panel 7), which could be related to claims of Mozart's greater exposition-recapitulation similarity. 

Features are present in all categories (other than the exposition), indicating the importance of features ranging from basic summary to sonata-inspired. 
This also suggests most statistical differences between Haydn's and Mozart's string quartets can be explained by the basic summary, interval, recapitulation, and development features. Of these categories, some are more commonly represented than others. For example, there are three interval features and three sonata-inspired features, while only one basic summary feature. The high count of interval features is expected, because of the fundamental role intervals serve in music. Similarly, the high count of sonata-inspired features reinforces the importance of the sonata style here, which is `central to the music of Haydn and Mozart' (Harutunian, 2005, p. 1).

The counts of features from each voice can describe distinctions between the composers' treatment of the instruments in the string quartet. There are two features from Violin 1, none from Violin 2, one from Cello, three from Viola, and one comparing both Cello and Viola.
These results suggest that Mozart and Haydn could be distinct in the way they compose the low accompanying voices, while their `leading' violin parts are more similar.
Although the string quartet genre has frequently been described as a `conversation', \cite{kim2018emergence} argues that the `democratic ideal' of four equal parts in a string quartet was rarely satisfied in the Classical period, due to first violin predominance.
In fact, Haydn often played the first violin part and his string quartets `... tend to lean on the first violinist as the main melody-bearer and virtuosic exhibitionist', while Mozart commonly played Viola and began to give `more weight ... to instruments other than the first violin'
(Kim, 2018, p. 27, 88). 
These potential differences in the composers' instrument weighing could explain the representation of mostly Viola and Cello features in our model. \cite{herlands2014machine} identified that Haydn has a higher rate of simultaneous onsets of rests between the three lower voices. While our feature selection methods did not identify that feature and involved a different dataset, Herlands et al. reached a similar conclusion to ours: Haydn uses the three lower voices `as a homogeneous accompanying unit', in contrast to Mozart's `varying groupings between the four voices' (p. 281). Additionally, Herlands et al. found that second intervals and other features were less associated with Haydn string quartets, reflecting `Haydn’s greater virtuosity in the first violin' (p. 281). Our `proportion of descending pairwise intervals for Violin 1' feature, which is generally higher for Haydn, could align with these claims of Haydn's first-violin virtuosity.

Features from pitch tracks outnumber features from duration tracks: there are six from pitch, while only two from duration. One explanation is that the role of pitch is more prominent than rhythm in classical Western music. 
Indeed, in a study with Western musical excerpts, \citet{schellenberg2000perceiving} found that pitch is more emotionally meaningful to listeners than rhythm. 

It is interesting that no exposition features were selected in the model, given the several musicological remarks regarding Haydn's more monothematic expositions relative to Mozart (\citealp{harutunian2005haydn, rosen1997classical}). 
Such findings could evidence that Mozart and Haydn string quartets differ more in their developments and recapitulations than their expositions. Another explanation is that our exposition features automatically search for repetitions of the first segment of notes in a piece, but some sonatas have introductions prior to presenting the main theme. Our recapitulation features, which also use the first segment of notes, could be affected too. Such complications reflect the challenges of automatic, algorithmic calculation of features, and follow-up work could improve the search of repeated themes, for both sonatas and pieces with other structures in the dataset.

\section{Conclusion}
We have conducted an interpretable machine learning study of Mozart versus Haydn string quartets, contributing quantitative results to the vast musicological scholarship of these composers. We proposed novel global features inspired by the sonata form that can be computed automatically from symbolic representations of musical scores. From these sonata-inspired features and other global features, the subset selection method of random ICM identified important musical features that distinguish between Haydn and Mozart string quartets. 
Our Bayesian logistic regression model containing these musical features achieved higher
classification accuracy than in previous studies: over $84\%$ for LOO on a benchmark dataset of 107 movements. For a dataset of 285 movements (almost all known movements), the same statistical methods yielded a LOO accuracy of about 82\%. Additional analyses on the larger dataset suggested seven features to be particularly important, including three sonata-inspired features. Many interpretations of these features aligned with claims previously made by musicologists, potentially offering quantitative validation of subjective musical information. These strong results indicate that Haydn and Mozart string quartets can be discriminated with high accuracy while offering musically relevant interpretations.  

Further directions for the Haydn-Mozart classification task could involve studying Mozart's No. 10 string quartets and Haydn's Op. 33 string quartets.
In 1781, Haydn claimed his Op. 33 string quartets were written ‘in an entirely new and special way’, and a number of scholars since then have attempted to characterise these distinctions \citep{klauk2016mozart}. Interestingly, there is some evidence that the Op. 33 string quartets `prompted Mozart’s return to this genre after almost ten years', since in 1785 Mozart published his No. 10 quartets (the `Haydn' quartets) with a dedication to Haydn (Klauk \& Kleinertz, 2016, p. 575). However, `the question of its [Op. 33's] relevance to Mozart's ``Haydn'' quartets has yet to be answered convincingly' (Klauk \& Kleinertz, 2016, p. 578). Quantitative analyses, such as through estimated probabilities or other similarity metrics, could offer new insights into this enduring historical and musicological debate. 

Follow-up research could extend Haydn-Mozart composer classification to similar tasks in MIR, such as music generation in the style of Mozart, Haydn, and their combination. If extended to more classical composers, estimated probabilities or other similarity metrics from composer classification could be applied to content-based artist recommendation, e.g., on websites like \url{https://musicroamer.com/}.
Additionally, the algorithms to calculate the sonata-inspired features in our study could be improved to more effectively detect sonata sections, repeated themes, and other Classical structures in these datasets.

Beyond Haydn and Mozart, our methods, novel sonata-inspired features, and their extensions could be used to classify similar classical composers, such as from the Classical period or writing in sonata, ternary, or related forms. Our strong predictive results provide preliminary evidence that structural information can improve symbolic-based classification of stylistically similar composers. Moreover, we view our work as an example of how interpretable machine learning can contribute to the field of MIR, so that compositions can be fully appreciated both musically and mathematically. As the mathematician James Joseph Sylvester (1908) famously wrote, `May not music be described as mathematics of the sense, mathematics as music of the reason?'.

\section*{Acknowledgements}
We would like to thank the reviewers for their careful and insightful comments, which greatly improved the work. We would also like to thank Dr. Peter van Kranenburg for sharing with us the HM107 dataset of Haydn and Mozart string quartet scores and Dr. Gissel Velarde for helping us access datasets for the study.

\bibliographystyle{apacite}
\bibliography{biblio}

\newpage

\beginsupplement
\section*{Supplementary Material}

All plots and analyses here can be found in the publicly accessible Github repository at \texttt{https://github.com/wongswk/haydn-mozart}.
Implementation of heatmaps are through the R package \textit{pheatmap} \citep{kolde2012pheatmap}. 

\bigskip
\noindent This supplement contains the following material, as referenced in the main text.
\begin{itemize}
    \item S1. Exploring Similarity among Segment Features (referenced in Section 3.2)
    \item S2. Investigating Dependence among Movements by Set, Quartet, and Date of Composition  (referenced in Section 5.2)
    \item S3. Comparing LOO on the Full Feature Set versus the Reduced Feature Set  (referenced in Section 5.2)
\end{itemize}

\section{Exploring Similarity among Segment Features}
As noted in Section 3.2, we have performed exploratory analyses to determine appropriate candidates for segment lengths of $m$ notes for the segment features. We compute all segment features for segment lengths $m = 8, 9, 10, \hdots, 17, 18$. For various subsets of these segment features, we then compute correlation matrices and visualize them via heatmaps. If $X$ is an $n \times p$ data matrix consisting of $n$ movements and $p$ standardized (centered and scaled) features, then the correlation matrix is computed as $X'X$; for $i, j = 1, 2, \hdots, p$, the entry $X'X_{ij}$ is the (Pearson sample) correlation between features $i$ and $j$ over the $n$ movements. From the heatmaps, we can discern that in most cases, features with segment lengths of distance 1 apart (e.g., segment lengths 8 and 9) have extremely high correlations of 0.9 or more. Meanwhile, when the segment lengths differ by 2, the correlations are weaker and in some cases lower than 0.5. Therefore, different information is captured for some features with segment lengths of distance 2 or more apart, while mostly the same information is reflected by segment lengths of distance 1 apart. Meanwhile, the number of proposed features would be increased by about 62\% (from 1182 to 1914) by including the segment features with segment lengths of distance 1 apart in the feature set. In particular, a dimensionality approaching 2000 features, relative to a sample size of 107 or 285, would potentially require more specialized feature selection or dimension reduction strategies. Thus, as a balance between maximizing statistical information and minimizing computational and modeling complexity, we find it reasonable to include only segment length $m = 8, 10, 12, 14, 16, 18$ as candidates for segment features in our feature set and use feature selection (Section 4) to more rigorously identify the optimal segment lengths.

\begin{figure}[H]
    \centering
    \includegraphics[width=.99\textwidth]{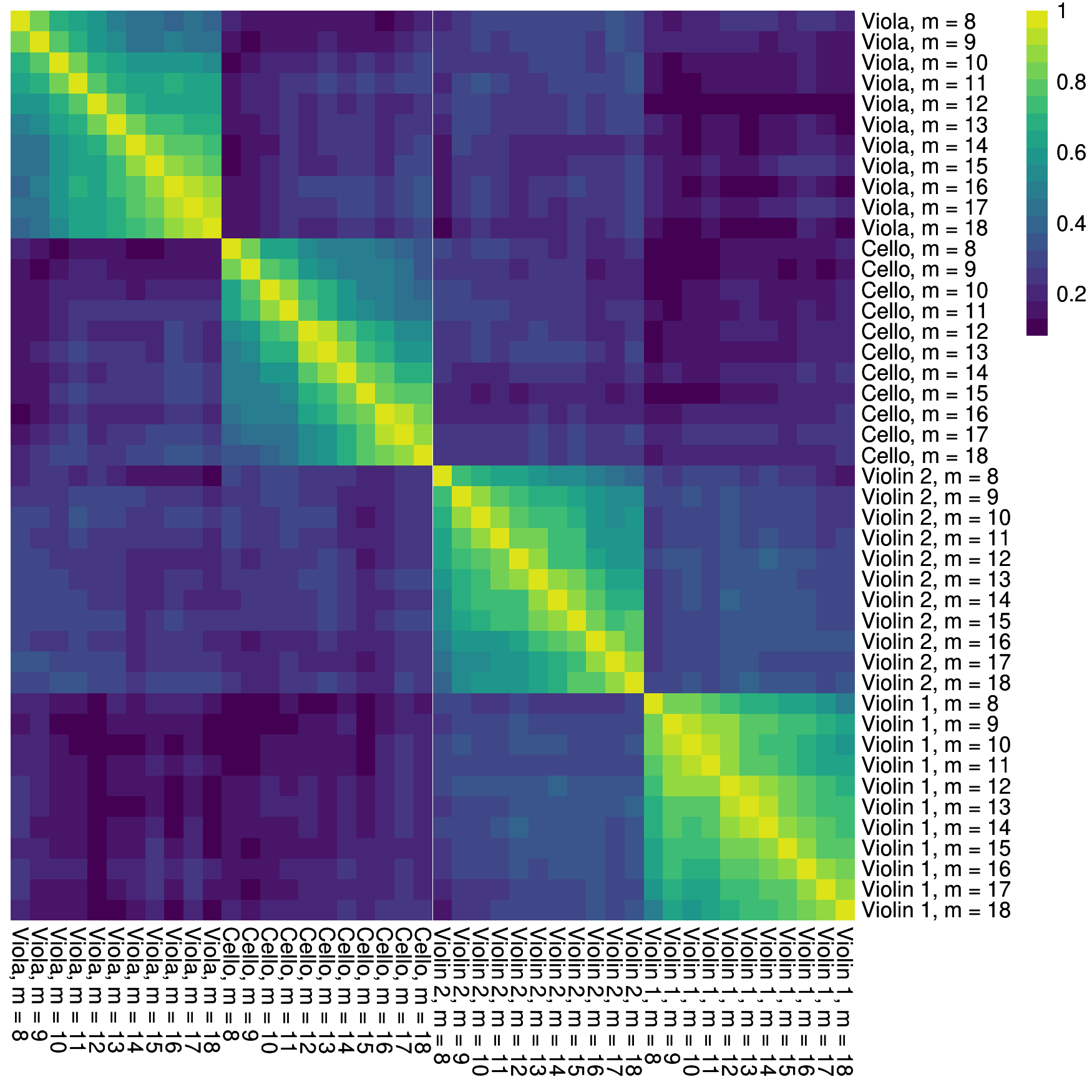}
    \caption{The heatmap shows the $p \times p$ correlation matrix $X'X$ for the recapitulation location duration features (for segment lengths of $m = 8, 9, 10, \hdots, 17, 18$ notes) in the HM285 dataset.}
    \label{fig:segment_heatmap}
\end{figure}

As an example, Figure \ref{fig:segment_heatmap} shows the correlation matrix for the recapitulation location features for duration in the HM285 dataset. The yellow diagonal represents correlations of 1, since each feature is perfectly correlated with itself. The four blocks forming around the diagonal correspond to each part of the string quartet, since, e.g., Viola features are correlated among themselves. Within each block, it is observed that correlations between pairs of segment features decrease as the distance between segment lengths increases. In particular, pairs of recapitulation segment features with segment lengths of distance 1 apart (e.g., $m = 8$ and $m = 9$) tend to have very strong correlations of 0.9 or higher;
pairs of segment features with segment lengths of distance 2 apart (e.g., $m = 8$ vs. $m = 10$) have slightly weaker correlations around 0.8 or lower. Heatmaps for the remaining subsets of features can be found in the Github repository. 

\section{Investigating Dependence among Movements by Set, Quartet, and Date of Composition}
As mentioned in Section 5.2, there are natural groupings among Haydn and Mozart string quartet movements that could potentially violate the assumption of independence required for validity of LOO CV. In addition to the potential within-quartet dependency, these composers often published quartets as sets, especially when commissioned, e.g., Haydn's `Prussian' quartets dedicated to King Frederick William II of Prussia or Mozart's No. 20 quartets (K. 499) published by Franz Anton Hoffmeister. More broadly, there could be temporal structure, since pieces are written over the course of a composer's life. 

We explore potential dependencies among movements by set, quartet, and date of composition through descriptive statistics: we visualize matrices of features and their correlations through heatmaps. 
Let $X = \begin{pmatrix} X_M \\ X_H \end{pmatrix}$ be the $n \times p$ data matrix consisting of $p$ standardized features (centered and scaled) for $n$ movements, ordered such that the first rows form a block $X_M$ of Mozart movements and the final rows form a block $X_H$ of Haydn movements. Within each composer block, we order each set of quartets by date of composition, then by quartet number, and finally by movement number.
For example, within the `Milanese' set of the Mozart block, the K. 155 quartets (movements 1, 2, then 3) would be followed by the K. 156 quartets. Define also the Gram matrix as $XX'$, which has $XX'_{ij}$ as the (Pearson sample) correlation between movement $i$ and movement $j$ over the $p$ standardized features, for all $i, j = 1, \hdots, n$.
We analyse the data matrix $X$, the Gram matrix $XX'$, and their variants, discussed as follows. 

\begin{figure}
    \centering
    \includegraphics{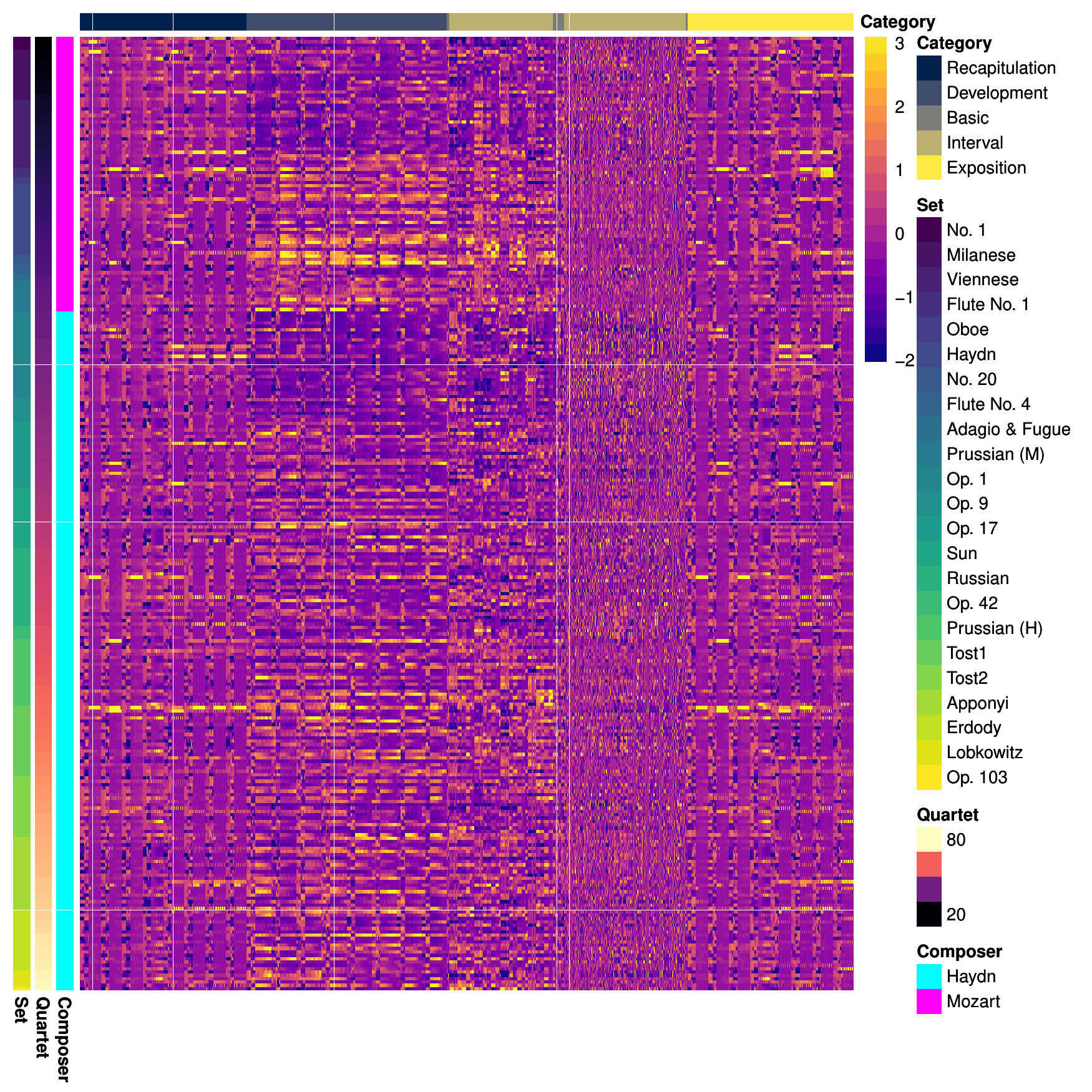}
    \caption{The heatmap represents the column-standardized, ordered $n \times p$ feature matrix $X$ on the HM285 dataset. The rows and columns correspond to movements and features, respectively. The left annotations mark set, quartet, and composer groups of each movement, while the top annotation denotes the category of each feature. }
    \label{fig:heatmap_X}
\end{figure}

\begin{figure}
    \centering
    \includegraphics{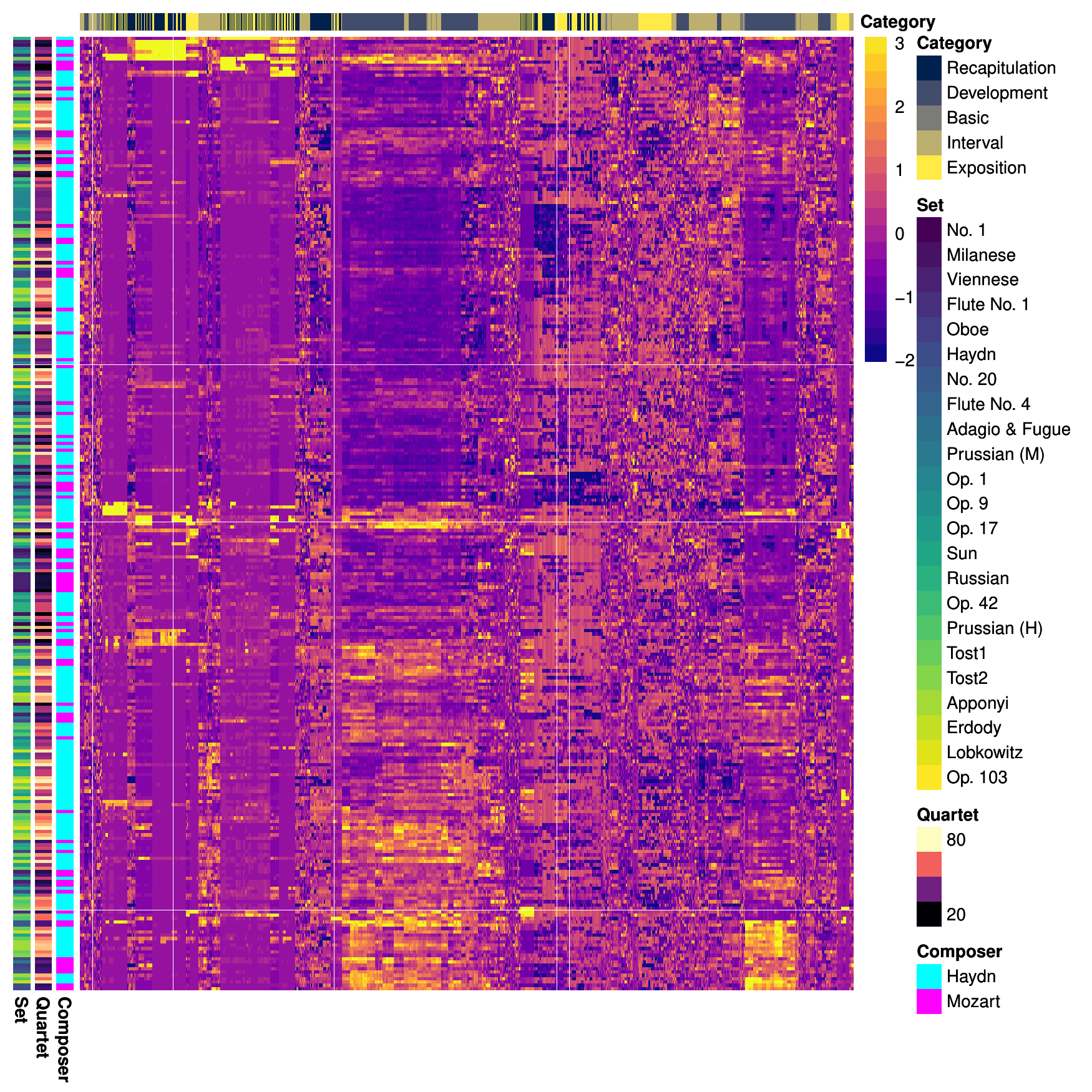}
    \caption{The heatmap represents the column-standardized $n \times p$ feature matrix $X$ on the HM285 dataset from Figure \ref{fig:heatmap_X}, with rows and columns reordered via hierarchical clustering.  The rows and columns correspond to movements and features, respectively. The left annotations mark set, quartet, and composer groups of each movement, while the top annotation denotes the category of each feature. }
    \label{fig:heatmap_cluster_X}
\end{figure}

The heatmap corresponding to the data matrix $X$ is displayed in Figure \ref{fig:heatmap_X}. Hierarchical clustering is then applied to $X$ to help identify potential patterns among movements and features. We opt for the standard choices of complete-linkage and Euclidean distance for clustering, resulting in the heatmap in Figure \ref{fig:heatmap_cluster_X}. If dependencies within sets, quartets, composer category, or date were present, the following patterns would be observed. In Figure \ref{fig:heatmap_X}, block patterns over the heatmap rows would form. In Figure \ref{fig:heatmap_cluster_X}, clusters corresponding to sets, quartets, composer, or date would form, which would be indicated through the column annotations. However, no such patterns emerge. We note that clustering does result in the formation of some other strong clusters (e.g., the bright yellow blocks at the bottom of the heatmap), but they appear unrelated to sets, quartet, date, or composer. This is an interesting result, because it suggests movements that could be musically similar, despite belonging to different groups. 
The same analysis is repeated for various subsets of movements and features, e.g., the `recapitulation features' and the `recapitulation features for the Mozart block $X_M$'; none of the resulting heatmaps (which can be found in the Github repository) show patterns by set, quartet, or date.

\begin{figure}
    \centering
    \includegraphics{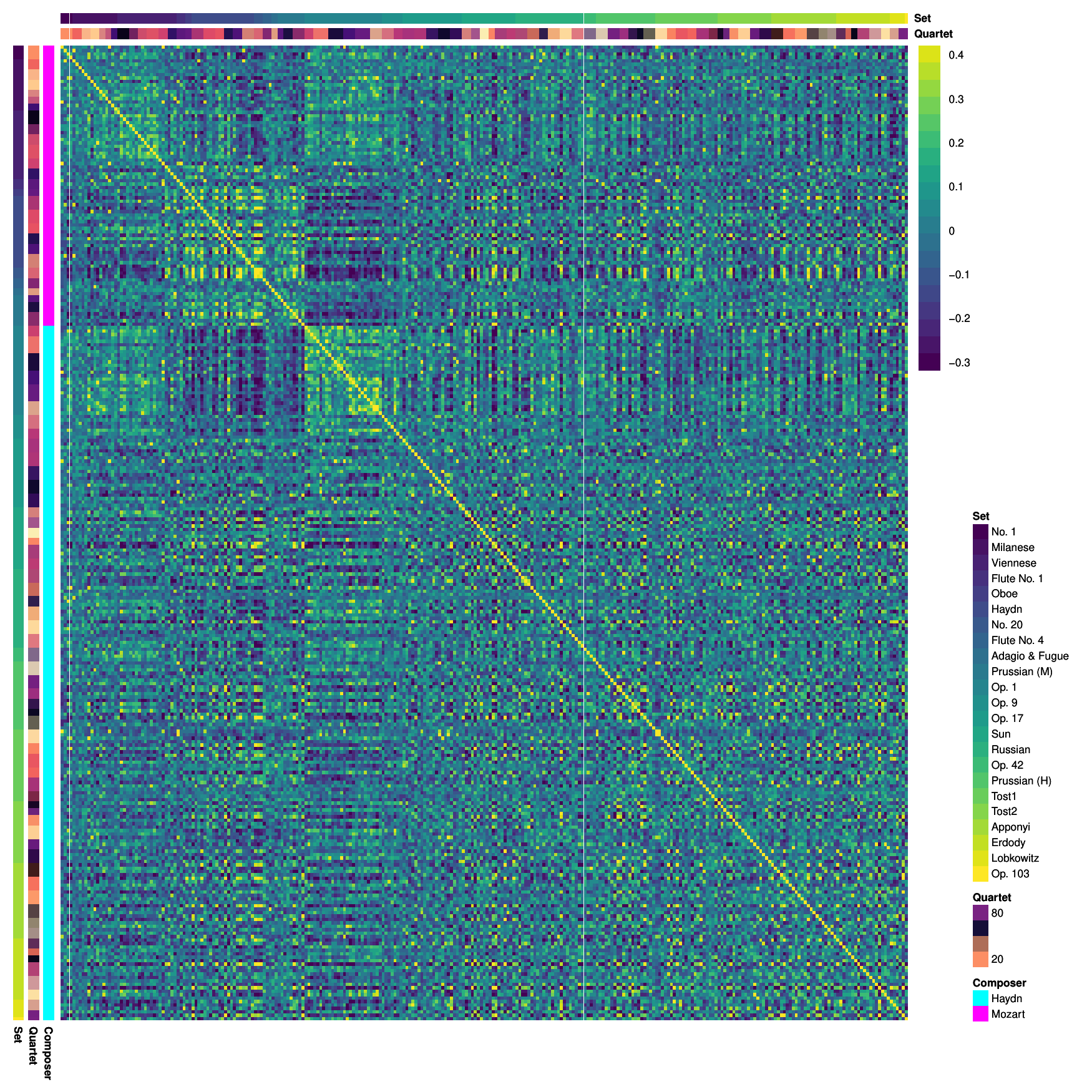}
    \caption{The heatmap depicts the $n \times n$ Gram matrix $XX'$ of the column-standardized, ordered feature matrix $X$. For all $i, j = 1, 2, \hdots, n$, row $i$ and column $j$ of $XX'$ corresponds to the correlation between rows $i$ and $j$ of $X$. The left annotations mark set, quartet, and composer groups of each movement, and the top annotations repeat the set and quartet groups for improved readability.}
    \label{fig:heatmap_Gram}
\end{figure}

The Gram matrix is visualized through the heatmap in Figure \ref{fig:heatmap_Gram}. Note that the colors associated with quartets have been randomized here (rather than following the order from Figures  \ref{fig:heatmap_X} and \ref{fig:heatmap_cluster_X}), for improved visualization. Similarly to the correlation matrix shown in Figure \ref{fig:segment_heatmap}, the yellow diagonal corresponds to correlations of 1 between each movement and itself.
Dependencies within sets, quartets, date, or composer would result in block-diagonal patterns on this heatmap; e.g., if the `Milanese' quartets had correlated features, then a square block of yellow colors would be visible along the diagonal. While we do observe a few blocks of yellow colors (e.g., in the upper left corner of the heatmap), the corresponding correlation values are quite weak at 0.2 to 0.4; additionally, their rectangular, non-square shapes indicate that they span multiple quartets and sets. Thus, any patterns in the Gram matrix heatmap are mild and seem more related to random chance than quartet, set, composer, or date. As above, we form heatmaps for other subsets of features and movements (included in the Github repository), but they display no notable patterns.

\begin{figure}
    \centering
    \includegraphics[width=.88\textwidth]{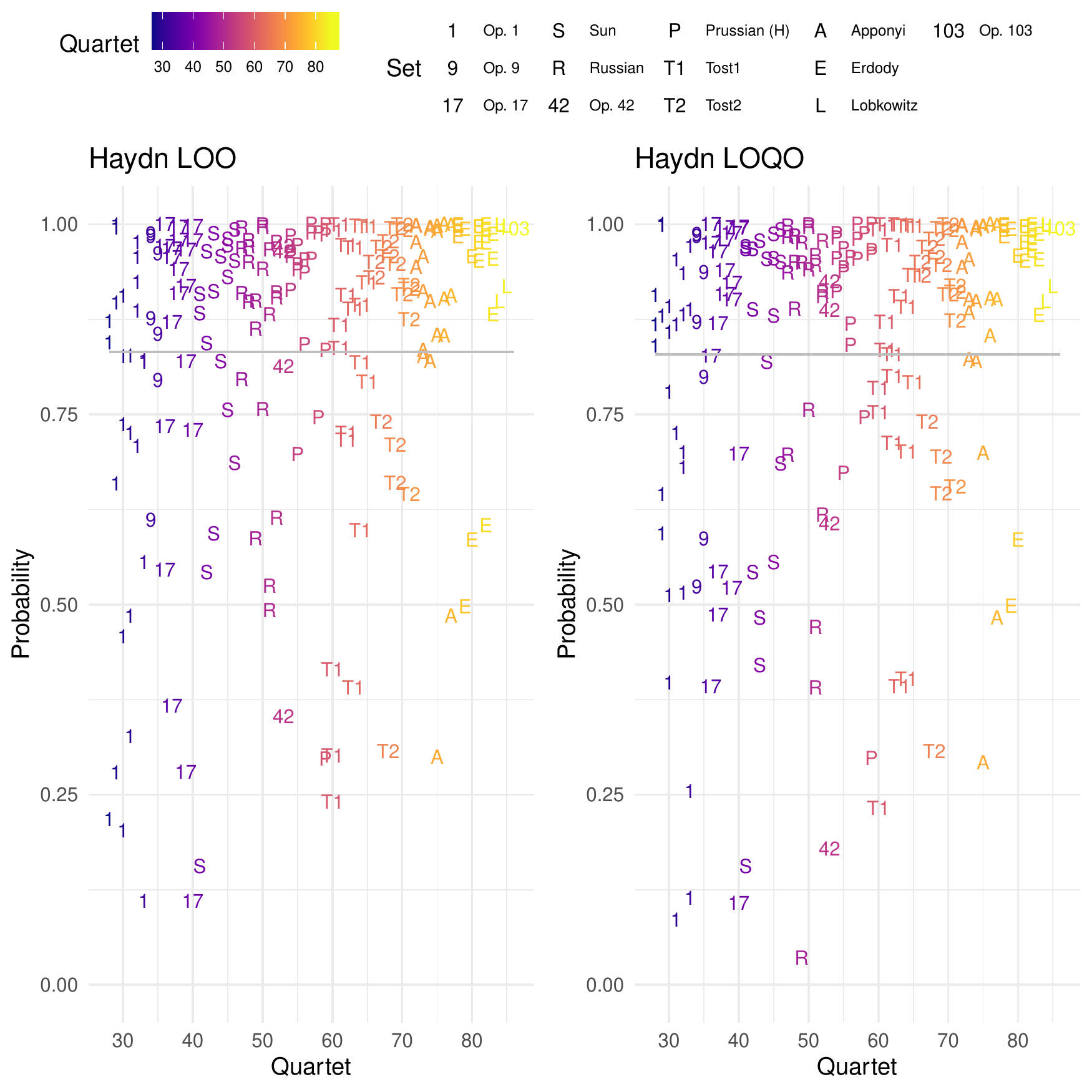}
    \caption{For Haydn movements in the HM285 dataset, we plot the estimated probabilities by quartet and set, which result from the LOO CV scheme (left) and the LOQO CV scheme (right). In each plot, the gray line marks the mean of the estimated probabilities over the movements. For improved visualization in these plots, quartets have been ordered by approximate date of composition then converted to a numerical scale.}
    \label{fig:probs_Haydn}
\end{figure}

\begin{figure}
    \centering
    \includegraphics[width=.88\textwidth]{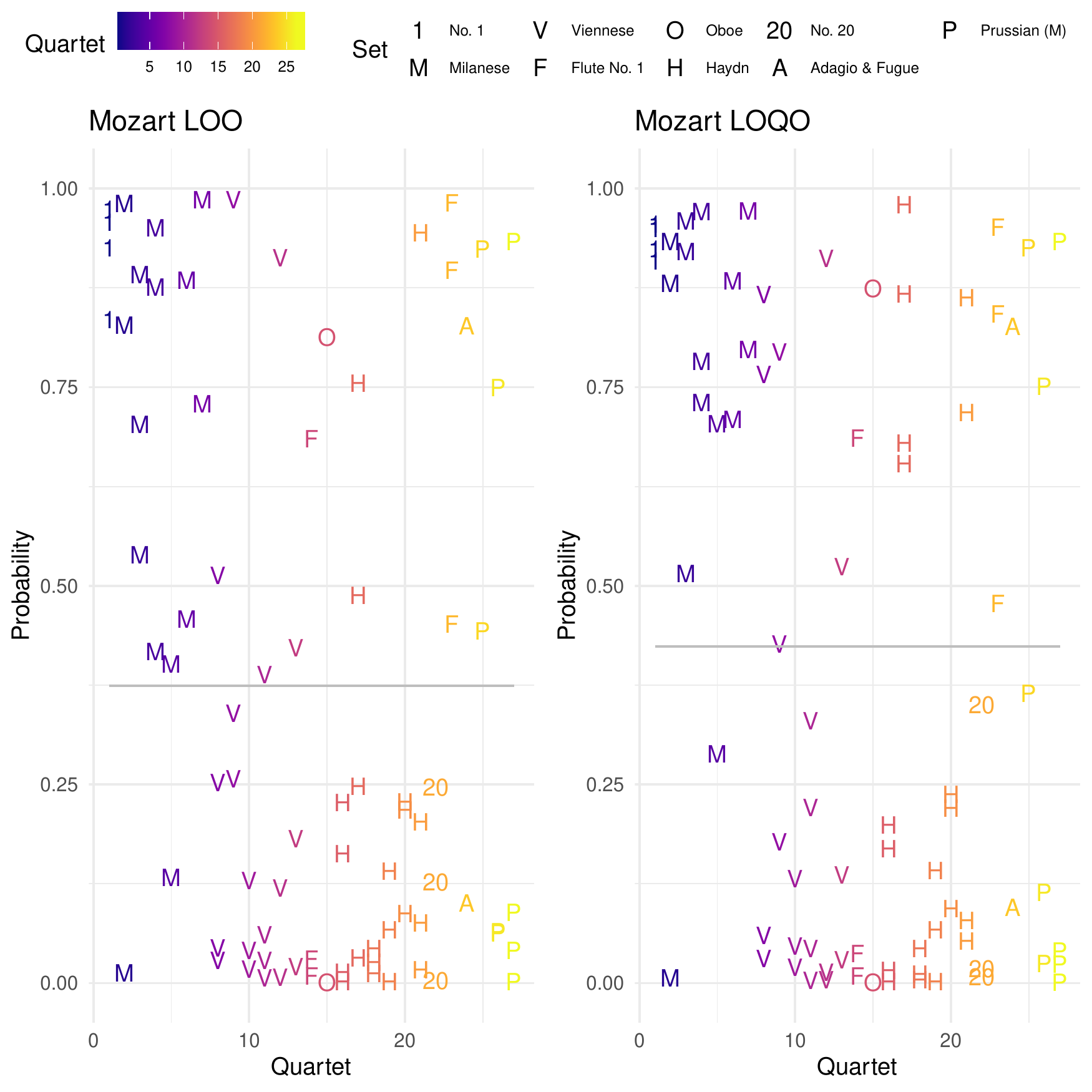}
    \caption{For Mozart movements in the HM285 dataset, we plot the estimated probabilities by quartet and set, which result from the LOO CV scheme (left) and the LOQO CV scheme (right). In each plot, the gray line marks the mean of the estimated probabilities over the movements. For improved visualization in these plots, quartets have been ordered by approximate date of composition then converted to a numerical scale.}
    \label{fig:probs_Mozart}
\end{figure}

Next, we analyse the estimated probabilities from LOO and LOQO to identify potential patterns. Figures \ref{fig:probs_Haydn} (Haydn) and \ref{fig:probs_Mozart} (Mozart) show estimated probabilities from LOO (left) and LOQO (right) by quartet and set. For each composer, the plots look very similar for LOO and LOQO, indicating that LOQO has not substantially affected the results, despite reducing the size of available training data in each CV fold. This stability across the CV schemes suggests that it was appropriate for movements within a quartet to be treated as approximately independent for LOO. Additionally, there are no discernible patterns by set or date of composition in these plots, implying that movements within a set or similar time period are also approximately independent.

Numeric summaries further demonstrate that the estimated probabilities and classes are similar for LOO and LOQO. As displayed in Table \ref{tab:LOQO_LOO} (top rows), close to half of the movements in HM285 (44.21\%) have the same estimated probabilities from LOO and LOQO CV, up to rounding. Any differences in estimated probabilities between the CV schemes infrequently translate to different predicted classes. Only about 9\% of movements in the dataset have different predictions for LOO versus LOQO, mostly corresponding to correct classifications in LOO and incorrect classifications in LOQO. While Mozart movements are more likely than Haydn movements to have different estimated probabilities between the two CV schemes (72\% for Mozart, 50\% for Haydn), Mozart is almost as likely as Haydn to have classifications that agree across CV (90\% for Mozart, 92\% for Haydn). Overall, these analyses confirm the stability and validity of our CV schemes. 

\begin{table}[]
    \centering
    \begin{tabular}{lll}
      \toprule
    Group    &Percentage& \\
    \midrule
    LOO probability $<$ LOQO probability &26.67\% \\
    LOO probability $=$ LOQO probability &44.21\% \\
    LOO probability $>$ LOQO probability &29.12\% \\
    \midrule 
    LOO class $<$ LOQO class &5.61\% \\
    LOO class $=$ LOQO class &91.93\% \\
    LOO class $>$ LOQO class &2.46\% \\
    \midrule 
    LOO probability $<$ LOO-reduced probability &45.96\% \\
    LOO probability $=$ LOO-reduced probability &2.11\% \\
    LOO probability $>$ LOO-reduced probability &51.93\% \\
    \midrule 
    LOO class $<$ LOO-reduced class &10.88\% \\
    LOO class $=$ LOO-reduced class &81.75\% \\
    LOO class $>$ LOO-reduced class &7.37\% \\
    \bottomrule
    \end{tabular}
    \caption{The estimated composer probabilities and corresponding classes from the Bayesian logistic regression model (with class label 0 for Mozart and class label 1 for Haydn) 
    are compared among LOO, LOQO, and LOO on the reduced feature set. Estimated probabilities for each movement have been rounded to the nearest hundredth when determining if they are equal here. }
    \label{tab:LOQO_LOO}
\end{table}

\section{Comparing LOO on the Full Feature Set versus the Reduced Feature Set}
As highlighted in Section 5.2, we compare the testing results of LOO on the full feature set versus the reduced feature set. Such stability comparisons can offer information about the contribution of our novel sonata-inspired features, which are excluded in the reduced feature set. We start by analysing the estimated probabilities and classes from each testing fold (left-out movement).
As displayed in Table \ref{tab:LOQO_LOO} (bottom rows), only about 2.11\% of the movements in HM285 have the same estimated probabilities (up to rounding) for LOO on the full set versus the reduced set. Recall from above that LOQO gave the same estimated probability as LOO for about half the movements.
About 18\% of movements in HM285 have different class predictions for LOO on the full versus reduced sets; in contrast, LOQO gave different class predictions than LOO for about 9\% of the movements. Additionally, we form plots that compare estimated probabilities on the full versus reduced feature sets, similarly as in Figures \ref{fig:probs_Haydn} and \ref{fig:probs_Mozart} and included in the Github repository.
Altogether, these graphical and numeric results show that probability and class stability is maintained across CV schemes (from LOO to LOQO), but instability occurs across feature sets (from the full feature set to the reduced feature set).

Next, we consider the stability with respect to selected features of LOO on the full feature set versus the reduced feature set. 
Of the 56 movements misclassified in LOO on the reduced subset, 34 of them were also misclassified in LOO on the full feature set, while 22 of them were classified correctly when given access to the full feature set. Examining the models for the latter 22 movements/folds can suggest potentially important features.
For LOO on the full feature set, all 22 of the models
contain at least one development feature (most commonly, the `standard deviation count' feature labeled (E) in Figure 3, represented in 20 models), and 20 of the models additionally contain a recapitulation feature (the `maximum fraction of overlap' feature labeled (G) in the figure). For LOO on the reduced feature set, each of the 22 models contain the basic `standard deviation of duration' feature (feature (A) in Figure 3), while 19 of the models include a `mean proportion of minor third intervals for Viola' feature (similar to feature (D) in the figure). These models imply that some of the basic and interval features can contribute to high composer classification accuracies, but greater success in discriminating Mozart requires the sonata-inspired features.

Interestingly, there are 16 movements misclassified from LOO on the full feature set but classified correctly from LOO on the reduced subset. One possibility is that these movements have structures outside sonata and related forms. However, there are no clear patterns among the movements, other than Haydn predominance (15 movements by Haydn and only 1 by Mozart). This result reinforces that, without the sonata-inspired features, a classifier tends to guess the majority class, Haydn.

\end{document}